\pdfoutput=1

\documentclass[twocolumn]{aastex62}
\usepackage{amsmath}
\usepackage{graphicx}
\usepackage{mathrsfs}
\usepackage{color}
\usepackage{url}
\usepackage{CJKutf8}
\usepackage{ulem}

\received{2020}
\revised{\today}
\accepted{\today}

\shorttitle{2I}
\shortauthors{Hui et al. 2020}

\begin{document}

\title{
Physical Characterisation of Interstellar Comet 2I/2019 Q4 (Borisov)
}

\correspondingauthor{Man-To Hui}
\email{manto@ifa.hawaii.edu}

\author{\begin{CJK}{UTF8}{bsmi}Man-To Hui (許文韜)\end{CJK}}
\affiliation{Institute for Astronomy, University of Hawai`i,
2680 Woodlawn Drive, Honolulu, HI 96822, USA}

\author{\begin{CJK}{UTF8}{bsmi}Quan-Zhi Ye (葉泉志)\end{CJK}}
\affiliation{Department of Astronomy,
University of Maryland,
College Park, MD 20742, USA}

\author{Dora F{\"o}hring}
\affiliation{Institute for Astronomy, University of Hawai`i,
2680 Woodlawn Drive, Honolulu, HI 96822, USA}


\author{Denise Hung}
\affiliation{Institute for Astronomy, University of Hawai`i,
2680 Woodlawn Drive, Honolulu, HI 96822, USA}

\author{David J. Tholen}
\affiliation{Institute for Astronomy, University of Hawai`i,
2680 Woodlawn Drive, Honolulu, HI 96822, USA}

\begin{abstract}

We present a study of interstellar comet 2I/2019 Q4 (Borisov) using both preperihelion and postperihelion observations spanning late September 2019 through late January 2020. The intrinsic brightness of the comet was observed to continuously decline throughout the timespan, likely due to the decreasing effective scattering cross-section as a result of volatile sublimation with a slope of $-0.43 \pm 0.02$ km$^{2}$ d$^{-1}$. We witnessed no significant change in the slightly reddish colour of the comet, with mean values of $\left \langle g - r \right \rangle = 0.68 \pm 0.04$, $\left \langle r - i \right \rangle = 0.23 \pm 0.03$, and the normalised reflectivity gradient across the {\it g} and {\it i} bands $\overline{S'} \left(g,i\right) = \left(10.6 \pm 1.4\right)$ \% per $10^3$ \AA, all unremarkable in the context of solar system comets. Using the available astrometric observations, we confidently detect the nongravitational acceleration of the comet following a shallow heliocentric distance dependency of $r_{\rm H}^{-1 \pm 1}$. Accordingly, we estimate that the nucleus is most likely $\la$0.4 km in radius, and that a fraction of $\ga$0.2\% of the total nucleus in mass has been eroded due to the sublimation activity since the earliest observation of the comet in 2018 December by the time of perihelion. Our morphology simulation suggests that the dust ejection speed increased from $\sim$4 m s$^{-1}$ in September 2019 to $\sim$7 m s$^{-1}$ around perihelion for the optically dominant dust grains of $\beta \sim 0.01$, and that the observable dust grains are no smaller than micron size.

\end{abstract}


\section{Introduction}

Cometary object 2I/2019 Q4 (Borisov) (formerly C/2019 Q4, hereafter ``2I") was discovered by G. Borisov on 2019 August 30 at apparent {\it R}-band magnitude $m_{R} \approx 18$ with a $\sim$7\arcsec~condensed coma.\footnote{See Minor Planet Electronic Circular 2019-R106 (\url{https://minorplanetcenter.net/mpec/K19/K19RA6.html}).} The orbital eccentricity of 2I is significantly hyperbolic ($e = 3.36$), indicating that 2I is unbound to the solar system and has an interstellar origin \citep{2019MNRAS.tmp.2747H}. Thus, 2I is the second interstellar small body ever observed in the solar system after 1I/2017 U1 (`Oumuamua) \citep{2018A&A...610L..11D}. As opposed to `Oumuamua, which appeared completely asteroidal in optical images by various observers \citep[e.g.,][]{2017ApJ...851L..38B,2017ApJ...850L..36J,2017ApJ...851L..31K}, 2I has been exhibiting an obvious cometary feature, indistinguishable from ordinary comets in the solar system in terms of its morphology and colour from the earliest observations since the discovery \citep{2019ApJ...885L...9F, 2019NatAs.tmp..467G, 2019ApJ...886L..29J, 2019A&A...631L...8O}. Therefore 2I is observationally the first known interstellar comet that visits the solar system. Remarkably, \citet{2020AJ....159...77Y} successfully identified 2I in prediscovery data from the Zwicky Transient Facility (ZTF) all the way back to mid-December 2018, when the object was $\sim$8 au from the Sun. Furthermore, 2I appears to be chemically distinct from majority of the known solar system comets, as the interstellar interloper was observed to contain substantially more carbon monoxide (CO) than water (H$_{2}$O) gas than any previously measured comets in the inner solar system \citep{2020NatAs.tmp...85B, 2020NatAs.tmp...84C}, possibly indicative of chemical composition of the protoplanetary disc where 2I is from appreciably different from our own.

In order to understand how 2I would evolve as it approached to the Sun and constraints on the physical characteristics of the object in contrast to typical solar system comets, we monitored 2I from late September 2019 to late January 2020, covering an arc from over two months prior to the perihelion passage ($t_{\rm p} = $ TDB 2019 December 8.6) to almost two months postperihelion. The paper is structured in the following manner. We describe the observations in Section \ref{sec_obs}, give results and analyses in Section \ref{sec_anls}, present discussions in Section \ref{sec_disc}, and conclude in Section \ref{sec_sum}.

\begin{figure*}
\gridline{\fig{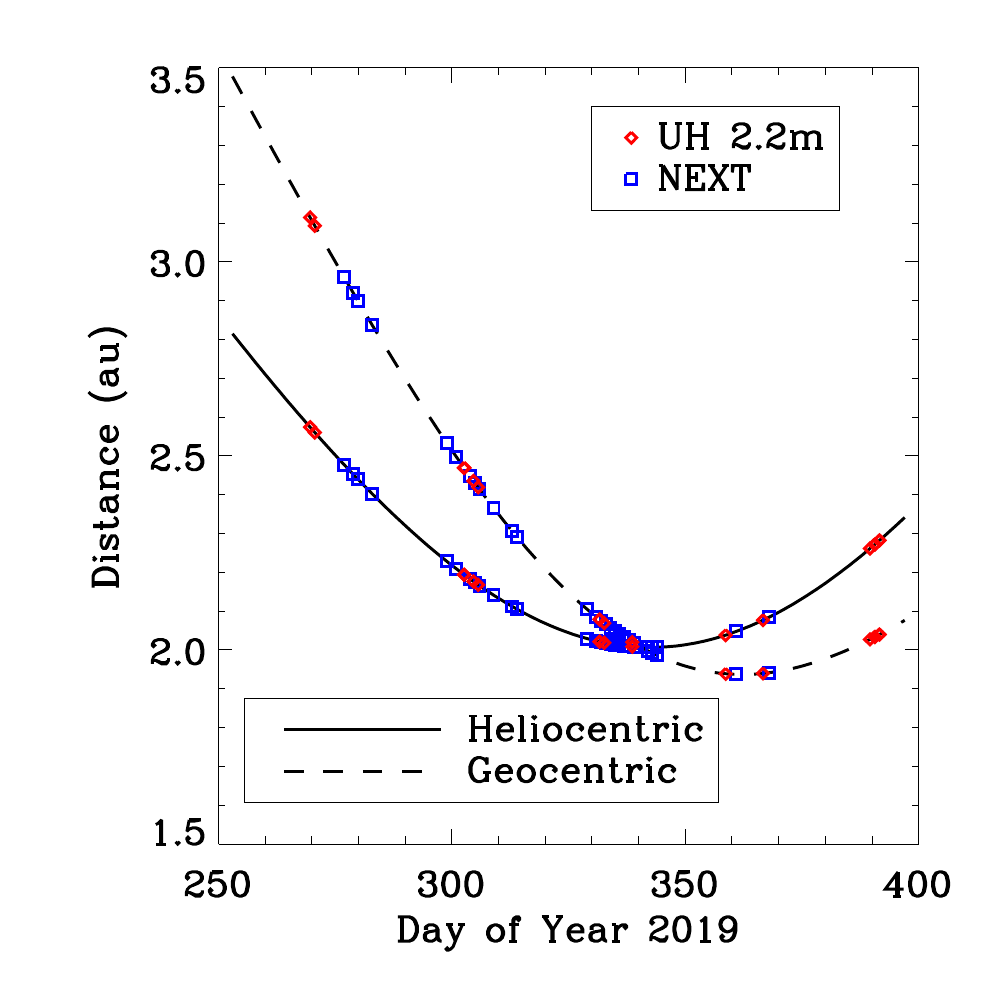}{0.5\textwidth}{(a)}
          \fig{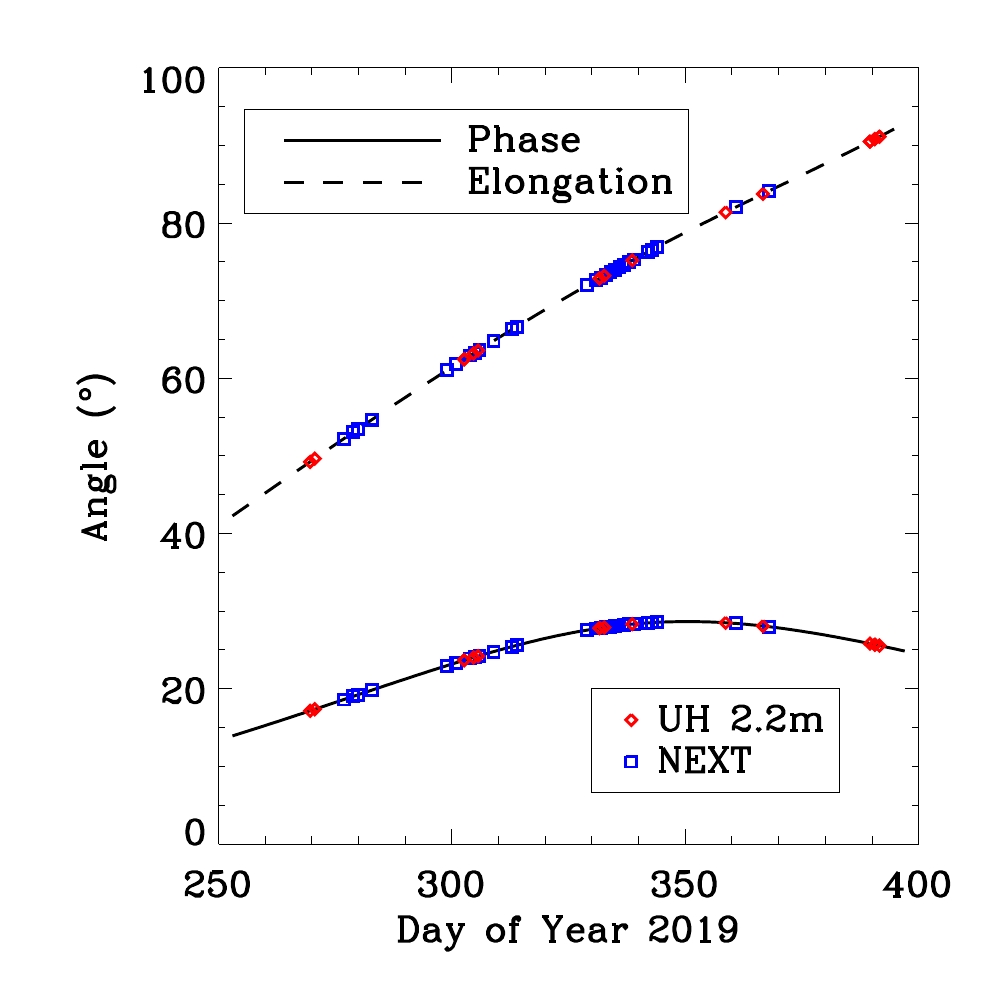}{0.5\textwidth}{(b)}
          }
\gridline{\fig{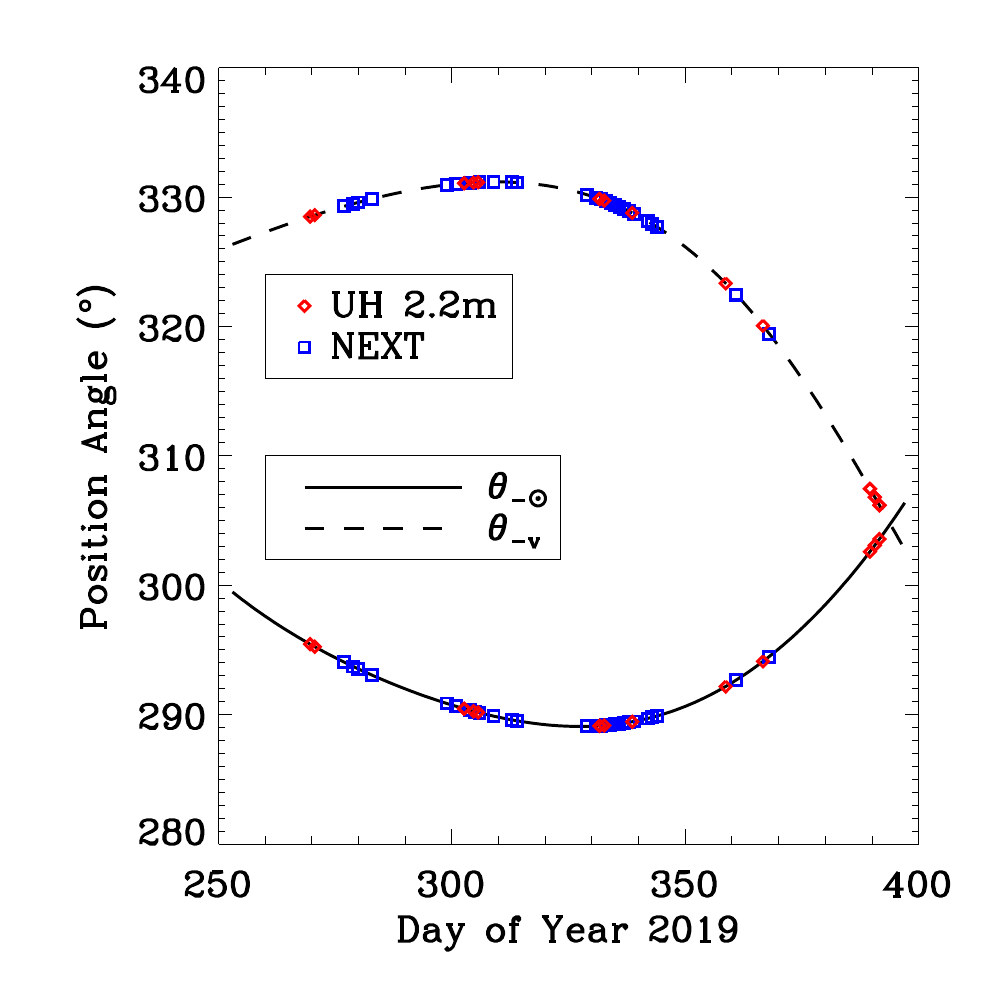}{0.5\textwidth}{(c)}
          \fig{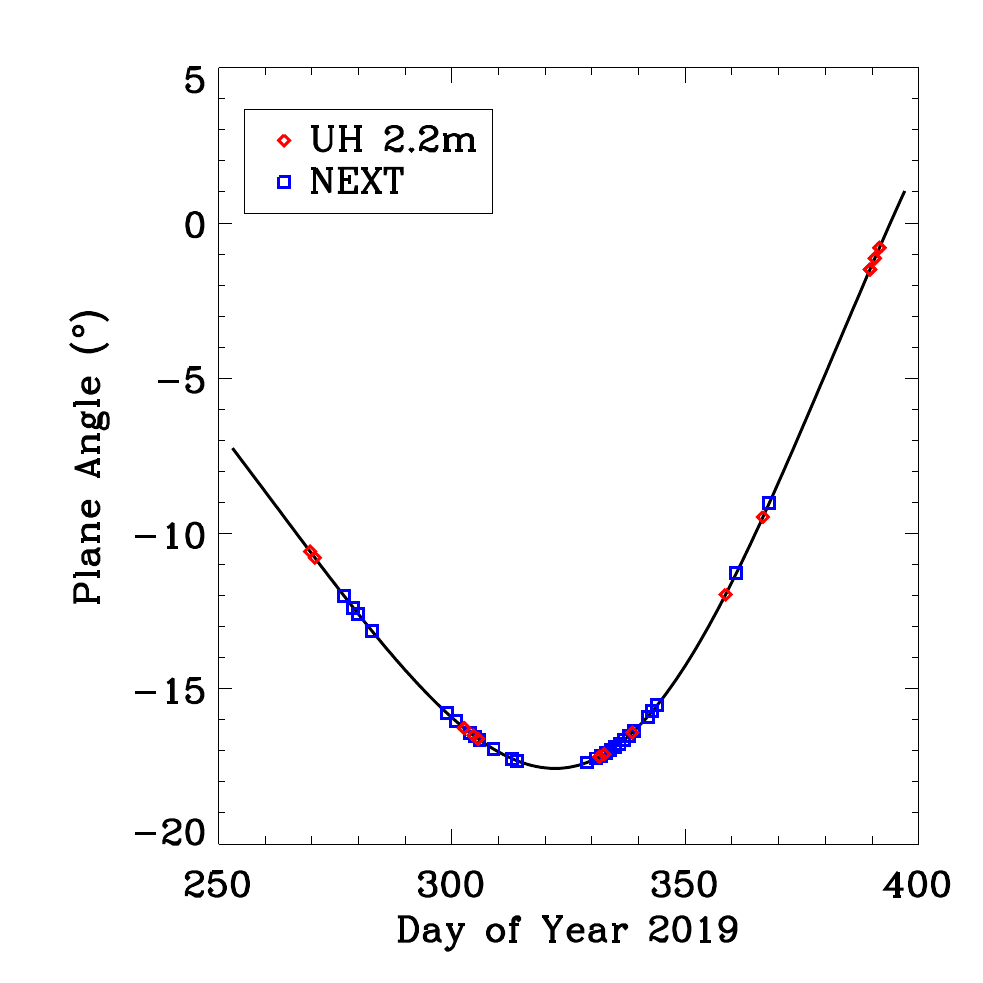}{0.5\textwidth}{(d)}
          }
\caption{
Observing geometry of interstellar comet 2I/2019 Q4 (Borisov) in terms of (a) heliocentric and geocentric distances, (b) phase angle and solar elongation, (c) position angles of projected antisolar direction ($\theta_{-\odot}$) and negative heliocentric velocity ($\theta_{-\bf{v}}$), and (d) the plane angle of the comet as functions of time during our observing campaign from the UH 2.2 m telescope (red diamonds) and NEXT (blue squares). 
\label{fig:2I_vgeo}
} 
\end{figure*}

\begin{figure*}
\epsscale{1.0}
\begin{center}
\plotone{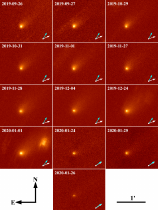}
\caption{
Median coadded images from the UH 2.2 m telescope of interstellar comet 2I/2019 Q4 (Borisov). Except that the two images from 2019 December 24 and 2020 January 01 are unfiltered due to the filter wheel issue, the others are in the {\it r} band. The trails in some of the panels are uncleaned artefacts from bright background stars. As indicated by the compass in the lower left, equatorial north is up and east is left. A scale bar of 1\arcmin~in length is shown. Also labelled are the position angles of the antisolar direction (white arrow) and the negative heliocentric velocity projected onto the sky plane (cyan arrow). Note that in late January 2020, as Earth was to cross the orbital plane of the comet, the two arrows become increasingly overlapped.
\label{img:2I_UH88}
} 
\end{center} 
\end{figure*}

\begin{figure}
\epsscale{1.1}
\begin{center}
\plotone{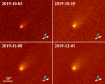}
\caption{
Selected examples of median coadded images from the 0.6 m NEXT telescope of interstellar comet 2I/2019 Q4 (Borisov). Only the image from 2019 October 03 is the {\it R} band, whereas the rest are in the {\it r} band after a renovation at Xingming Observatory. All images have the equatorial north up and east left. A scale bar of 30\arcsec~in length is shown. Same as in Figure \ref{img:2I_UH88}, the position angles of the antisolar direction (white arrow) and the negative heliocentric velocity projected onto the sky plane (cyan arrow) are labelled.
\label{img:2I_NEXT}
} 
\end{center} 
\end{figure}

\section{Observations}
\label{sec_obs}

We conducted observations of 2I using the University of Hawaii 2.2 m telescope and a Tektronix $2048 \times 2048$ CCD camera at the f/10 Cassegrain focus, through {\it g}', {\it r}', and {\it I}-band filters.  To improve the temporal coverage of the comet, we included in our analysis publicly available
data from the 0.6-m NEXT telescope at Xingming Observatory located in Xinjiang, China. Images from the UH 2.2 m telescope were tracked nonsidereally following the apparent motion of the comet. Due to a mechanical failure of the camera's filter wheel in December 2019 and the fact that the primary observations on these nights were made in white light, only unfiltered images of the comet were taken on 2019 December 24 and 2020 January 01. The images have a square field-of-view (FOV) of $7\farcm5 \times 7\farcm5$, and were $2 \times 2$ binned on chip, resulting in an image scale of 0\farcs44 pixel$^{-1}$. In order to mitigate artefacts such as cosmic ray hits, bad CCD columns and dead pixels, we dithered images between each exposure. Seeing during these observations varied between $\sim$0\farcs7 and 1\farcs0, typically $\sim$0\farcs8 (full width at half maximum, or FWHM, of field stars). 

We also included data acquired from NEXT to improve the temporal coverage of the comet. Initially the images were taken through {\it BVRI} filters, and were later switched to the Sloan {\it gri} system starting from early October 2019 after a renovation of the observatory. Since the telescope could only follow the comet in a sidereal rate, an individual exposure time of 120 s was set so as to keep the trailing of the comet visually unnoticeable. The images have an image scale of 0\farcs63 pixel$^{-1}$ in the $1 \times 1$ binning mode, with a FOV of $21\farcm5 \times 21\farcm5$. Seeing at NEXT, typically $\sim$2\arcsec-3\arcsec, was incomparable to that at the UH 2.2 m telescope, owing to a much lower elevation of the observatory.

All of the images were calibrated in a standard fashion, i.e., subtracted by bias frames taken from each night, and divided by flat-field frames that were generated from the science images in the same filters from the same nights, or neighbouring nights in a few cases, to fully eradicate influences from field stars and the comet. An additional step for the NEXT data was that before flatfielding dark frames were subtracted from the images. Cosmic rays and bad pixels were removed by L.A.Cosmic \citep{2001PASP..113.1420V} and the IRAF task {\tt cosmicrays}.

We show the observing geometry of 2I from the two telescopes in Figure \ref{fig:2I_vgeo}.

\begin{figure*}
\gridline{\fig{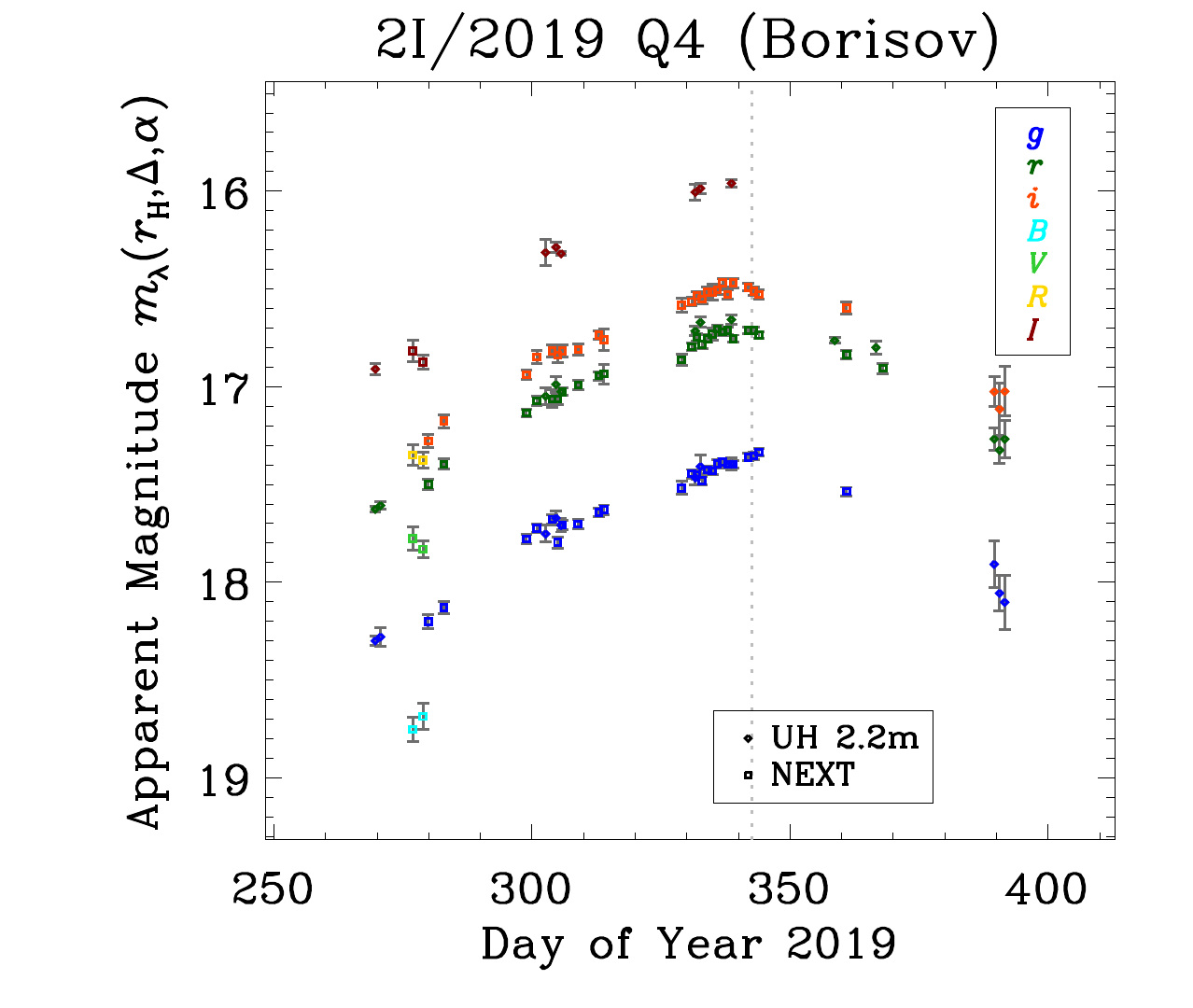}{0.5\textwidth}{(a)}
          \fig{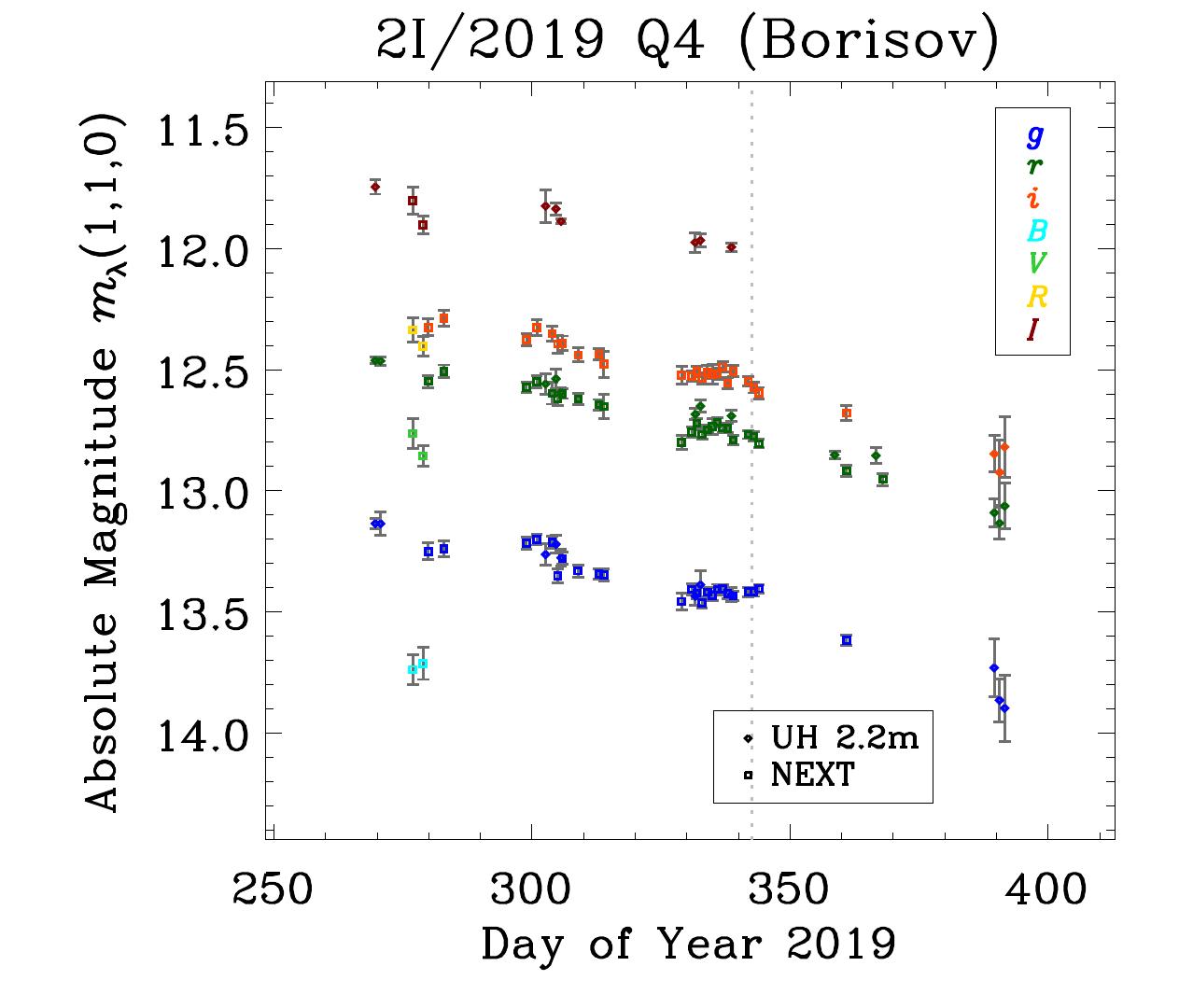}{0.5\textwidth}{(b)}
          }
\caption{
The apparent (a) and intrinsic (b) lightcurves of interstellar comet 2I/2019 Q4 (Borisov) as functions of time during our observing campaign from the UH 2.2 m telescope (diamonds) and NEXT (squares). The reduction bands are colour coded as indicated in the legend. Panel (b) was obtained by applying Equation (\ref{eq_m_abs}) to normalise the apparent magnitude of the comet in panel (a) to $r_{\mathrm{H}} = \mathit{\Delta} = 1$ au and $\alpha = 0$\degr. Assuming that 2I has a phase function similar to those of the solar system comets, thereby approximated by the empirical Halley-Marcus function by \citet{2007ICQ....29...39M} and \citet{2011AJ....141..177S}, we can note in panel (b) that its intrinsic brightness has been fading since our earliest observation from late September 2019. The vertical grey dotted line in each of the panels marks the perihelion epoch of the comet ($t_{\rm p} = $ TDB 2019 December 8.6).
\label{fig:2I_lc}
} 
\end{figure*}

\section{Analysis}
\label{sec_anls}

\subsection{Photometry}
\label{ss_phot}

Photometric measurements were performed slightly differently on images from the two telescopes. For data taken from the UH 2.2 m telescope, we measured the flux of 2I in the individual images, whereas for data from NEXT, we measured the flux on nightly median combined images through the same filters with alignment on the apparent motion of the comet. The aperture has a fixed projected linear radius of $\varrho = 10^4$ km at the topocentric distance of the comet so as to minimise potential biases from the aperture effect. The angular size of the chosen photometric aperture ($\ga$4\farcs6 in radius) is always larger than the seeing FWHM while the signal-to-noise ratio (SNR) is close to maximal. The sky background was computed from a concentric annulus with inner and outer radii of 3$\times$ and 6$\times$ the photometric aperture radius. We have made tests by changing the annulus size, but the results are always consistent within uncertainties, which were determined from Poisson statistics of the CCD. We have also repeated the measurements with a fixed aperture of $\varrho = 1.5 \times 10^4$ km in radius. The general shape of the lightcurve of the comet is the same, but with larger uncertainties and visually slightly greater scatter due to decrease in the SNR of the comet. So we conclude that our results should be robust. 

On a few occasions for the UH 2.2 m observations taken prior to late January 2020, there are faint field stars that partly fall within the photometric aperture in some of the images. To clean the contamination, we first extracted a number of field stars from the same individual images using StarFinder \citep{2000A&AS..147..335D}, then fitted and obtained trailed PSF models, which were scaled by brightness and subtracted from the images. These stars have all been cleaned nicely, leaving no noticeable artefacts in the photometric aperture. In late January 2020, the comet was near the galactic equator and therefore the FOVs are packed with stars. We computed nightly median images registered on field stars in respective filters, in which the comet was removed nicely. These median images were used as templates for optimal subtraction with High Order Transform of PSF ANd Template Subtraction \citep[HOTPANTS;][]{2015ascl.soft04004B}, resulting in a much cleaner sky background. For the NEXT images, contamination of faint field stars is not a concern because they were removed in the nightly median combined stacks.

Image zeropoints of the UH 2.2 m telescope were obtained from photometry of field stars in the individual images using an aperture of 9 pixels ($\sim$4\farcs0) in radius, whereas for the NEXT data, we measured the image zeropoints on nightly median combined stacks with alignment of field stars using an aperture of 10 pixels ($\sim$6\farcs3) in radius. Such apertures are large enough to enclose the majority of flux of the field stars and avoid aperture corrections due to varying seeing. Sky backgrounds were measured with a sky annulus having inner and outer radii 1.5$\times$ and 2.5$\times$ the corresponding aperture radii. Magnitudes of the field stars were taken from the Pan-STARRS1 (PS1) Data Release 1\citep[DR1;][]{2016arXiv161205243F}, and were transformed from the PS1 system to the corresponding photometric bands using equations by \citep{2012ApJ...750...99T} for all of the observations but those from 2019 December 26 and thereafter, when the comet was at decl. $<-30$\degr. In these cases we switched to the AAVSO Photometric All-Sky Survey Data Release 10 \citep[APASS DR10;][]{2018AAS...23222306H}.\footnote{Accessible at \url{https://www.aavso.org/apass-dr10-download}.} As for the unfiltered observations from the UH 2.2 m telescope, we initially included a linear colour slope to colour index {\it g} $-$ {\it r}. However, after tests we immediately discovered that the colour slope is statistically zero at the $1\sigma$ level determined from field stars having colour indices in a range of $0.3 \le g - r \le 1.0$, and therefore we ignored the colour term in the final version of the photometric reduction.

No significant systematic trend in the lightcurve of 2I is seen over the timespans (all lasted $<$1 hr) from our observing nights from the UH 2.2 m telescope, likely due to coma dilution \citep{1991ASSL..167...19J}. Thus, we only present the weighted mean apparent magnitude of 2I and the corresponding standard deviation of the repeated measurements in the same filters from the same nights in Figure \ref{fig:2I_lc}a, together with measurements from NEXT. 

The observing geometry of the comet varied considerably during our observation campaign, and thus must be corrected so as to investigate the intrinsic brightness of the comet from the apparent magnitude of 2I, denoted as $m_{\lambda} \left(r_{\rm H}, {\it \Delta}, \alpha \right)$, where $r_{\rm H}$ and ${\it \Delta}$ are respectively the heliocentric and topocentric distances of the comet, and $\alpha$ is the phase angle, via the following equation:
\begin{equation}
m_{\lambda} \left(1,1,0\right) = \underbrace{m_{\lambda} \left(r_{\rm H}, {\it \Delta}, \alpha \right) - 5 \log \left(r_{\rm H} {\it \Delta}\right)}_{m_{\lambda} \left(1, 1, \alpha \right)}  + 2.5 \log \phi \left(\alpha \right).
\label{eq_m_abs}
\end{equation} 
\noindent Here, $m_{\lambda} \left(1, 1, \alpha \right)$ is termed the reduced magnitude, and $\phi$ is the phase function of the comet, which is assumed to resemble those of the solar system comets and approximated as the empirical Halley-Marcus phase function by \citet{2007ICQ....29...39M} and \citet{2011AJ....141..177S} (see Section \ref{ss_phi} for further discussions of the phase-angle correction). We present the results in Figure \ref{fig:2I_lc}b. We thus notice that, although the apparent brightness of 2I was increasing on its way to perihelion and started to fade afterwards, the intrinsic brightness in fact has been steadily decreasing since our earliest observation of the comet from the UH 2.2 m telescope in September 2019. This result appears to contradict the earliest observations of the comet that cover much shorter timespans \citep{2019ApJ...886L..29J,2020ApJ...888L..23J}.

\subsection{Nongravitational Acceleration}
\label{ss_NG}

\begin{deluxetable*}{lr|c|c|c|c}
\tablecaption{Nongravitational Parameters of Comet 2I/2019 Q4 (Borisov)
\label{tab:NG}}
\tablewidth{0pt}
\tablehead{
\multicolumn{2}{c|}{Nongravitational Force Model} &
\multicolumn{3}{c|}{Nongravitational Parameters (au d$^{-2}$)} & Mean Residuals \\  \cline{3-5}
\multicolumn{2}{c|}{} & Radial $A_1$ & Transverse $A_2$ & Normal $A_3$ & (\arcsec)
}
\startdata
$\sim r_{\rm H}^{-n}$ & Power-law index $n = -1$
                  & $\left(+2.54 \pm 0.14 \right) \times10^{-9}$
                  & $\left(-1.27 \pm 2.17 \right) \times10^{-10}$
                  & $\left(-6.00 \pm 1.08 \right) \times10^{-10}$
                  & $\pm 0.88830$ \\
 & $0$
                  & $\left(+4.52 \pm 0.41 \right) \times10^{-9}$
                  & $\left(-1.69 \pm 5.66 \right) \times10^{-10}$
                  & $\left(-2.09 \pm 0.39 \right) \times10^{-9}$
                  & $\pm 0.88790$ \\
 & $+1$
                  & $\left(+7.12 \pm 1.08 \right) \times10^{-9}$
                  & $\left(-1.33 \pm 1.26 \right) \times10^{-9}$
                  & $\left(-6.66 \pm 1.03 \right) \times10^{-9}$
                  & $\pm 0.88781$ \\ 
 & $+2$
                  & $\left(+1.19 \pm 0.24 \right) \times10^{-8}$
                  & $\left(-2.78 \pm 2.49 \right) \times10^{-9}$
                  & $\left(-1.63 \pm 0.22 \right) \times10^{-8}$
                  & $\pm 0.88796$ \\ 
 & $+3$
                  & $\left(+2.26 \pm 0.49 \right) \times10^{-8}$
                  & $\left(-3.28 \pm 4.86 \right) \times10^{-9}$
                  & $\left(-3.53 \pm 0.44 \right) \times10^{-8}$
                  & $\pm 0.88832$ \\ 
 & $+4$
                  & $\left(+4.55 \pm 0.99 \right) \times10^{-8}$
                  & $\left(+1.08 \pm 9.69 \right) \times10^{-9}$
                  & $\left(-7.35 \pm 0.88 \right) \times10^{-8}$
                  & $\pm 0.88874$ \\ 
 & $+5$
                  & $\left(+9.33 \pm 2.00 \right) \times10^{-8}$
                  & $\left(+8.37 \pm 19.66 \right) \times10^{-9}$
                  & $\left(-1.51 \pm 0.18 \right) \times10^{-7}$
                  & $\pm 0.88914$ \\ \hline
\multicolumn{2}{l|}{H$_{2}$O-ice sublimation\tablenotemark{$\dagger$}}
                  & $\left(+2.80 \pm 0.58 \right) \times10^{-8}$
                  & $\left(+7.39 \pm 5.83 \right) \times10^{-9}$
                  & $\left(-4.51 \pm 0.51 \right) \times10^{-8}$
                  & $\pm 0.88998$ \\
\enddata
\tablenotetext{\dagger}{The isothermal model by \citet{1973AJ.....78..211M}.}
\tablecomments{
We included the same 2842 out of 2994 in total astrometric observations from 2018 December 13 to 2020 February 27 to obtain the solutions for all of the models. Observations with residuals in excess of 3\arcsec~were regarded as outliers. See Section \ref{ss_NG} for detailed information.}
\end{deluxetable*}

We obtained astrometry of 2I in the {\it r}-band images and the unfiltered ones from the nights when the filter wheel malfunctioned at the UH 2.2 m telescope with reduction using the Gaia DR2 catalogue \citep{2018A&A...616A...1G}. No astrometric measurement was done with the NEXT images because of the low SNR of the comet and much worse seeing. Astrometric observations from other stations, downloaded from the Minor Planet Center (MPC) Observation Database\footnote{\url{https://minorplanetcenter.net/db_search}.}, in combination with our measurements, were then fed into orbit determination code FindOrb\footnote{\url{https://www.projectpluto.com/find_orb.htm}}. We debiased the astrometry by following the method described in \citet{2015Icar..245...94F}, followed by assignment of a weighting scheme detailed in \citet{2017Icar..296..139V} for observations downloaded from the MPC. This procedure was used because many observations are reported without positional uncertainties, and in cases for which they were reported, the MPC is not yet exporting them. Our astrometric measurements and those from the ZTF, including the prediscovery observations by \citet{2020AJ....159...77Y}, were weighted by the measured positional uncertainties. In addition to the gravitational force by the Sun, perturbations from the eight major planets, Pluto, the Moon, and the most massive 16 main-belt asteroids, and the relativistic corrections were taken into account, although these were found to be unimportant in comparison to the gravitational effect from the Sun. The planetary and lunar ephemerides DE 431 \citep{2014IPNPR.196C...1F} were utilised.

Initially we attempted to determine a gravity-only orbit solution to the astrometric observations of 2I. However, we soon found that there exists a strong systematic trend in the astrometric residuals that could not be removed no matter how we adjusted the residual cutoff threshold. For example, the majority of our astrometric measurements from late January 2020 would have astrometric residuals $\ga 5\sigma$, with one even exhibiting a deviation at $\sim11\sigma$, whereas the ZTF prediscovery positions are deviated from the calculated counterparts by $\ga 5\sigma$. The systematic trend remains even if we opt to discard the ZTF prediscovery observations. Therefore, we decided to further include the radial, transverse, and normal (RTN) nongravitational parameters, corresponding to $A_{j}$ ($j=$1,2,3), which were first introduced by \citet{1973AJ.....78..211M} and now have been widely applied, as free parameters to be solved in FindOrb. Instead of directly adopting the nongravitational force model by \citet{1973AJ.....78..211M} based on isothermal sublimation of water ice for reasons described in \citet{2017AJ....153...80H}, we tested models scaled as $r_{\rm H}^{-n}$ and varied the power-law index $n$. Astrometry with residuals in excess of 3\arcsec~were rejected as outliers.\footnote{We have tested that even if we skip the step of outlier rejection, the resulting RTN nongravitational parameters are statistically similar to the ones with the step performed within the $1\sigma$ level.}

The obtained results, including those from the model by \citet{1973AJ.....78..211M} to form a comparison, are presented in Table \ref{tab:NG}. We clearly see that the nongravitational force models with heliocentric distance dependencies of $\sim r_{\rm H}^{-1 \pm 1}$ provide the best fits amongst the models we tested. Importantly, the strong systematic trend in the astrometric residuals no longer exists in the best fits. In particular, the astrometric residuals of our measurements and those from the ZTF are always at the $\la 1\sigma$ level. Had a steep heliocentric distance dependency for the nongravitational force model been more suitable, the magnitude of the out-of-plane nongravitational parameter $A_3$ will be unusually enormous in the context of solar system small bodies (referenced to the JPL Small-Body Database Search Engine). We thereby infer that the mass-loss rate of 2I most unlikely varies steeply with the heliocentric distance. In this regard, 2I is distinct from comet 67P/Churyumov-Gerasimenko, the target of the Rosetta mission, with a steeper heliocentric distance dependency for the mass-loss rate \citep[e.g., $\sim r_{\rm H}^{-5.6}$ for H$_{2}$O;][]{2016MNRAS.462S.156F, 2020Icar..33513421C}, and thereby a significantly steeper power-law nongravitational acceleration preferred by the fit to all of its astrometric data since 1995 (steeper than $\sim r_{\rm H}^{-4}$; D. Farnocchia, private communication).

We note that the fit with the isothermal H$_{2}$O-ice sublimation by \citet{1973AJ.....78..211M} is the worst compared to the models with $\sim r_{\rm H}^{-n}$ for $-1 \le n \le 5$ (Table \ref{tab:NG}). Although the astrometric residuals of the ZTF prediscovery observations are considerably smaller than those in the gravity-only solution, they are still $\ga2\arcsec$ and only half of them are at the $\la 1\sigma$ level. The poor fit strongly indicates that the nucleus of 2I is not subject to a nongravitational acceleration following the model by \citet{1973AJ.....78..211M} in which they assumed isothermal sublimation activity of water ice without a mantle layer. This find might also call into question the applicability of the model by \citet{1973AJ.....78..211M} to general solar system comets.

\section{Discussion}
\label{sec_disc}

\begin{figure}
\epsscale{1.2}
\begin{center}
\plotone{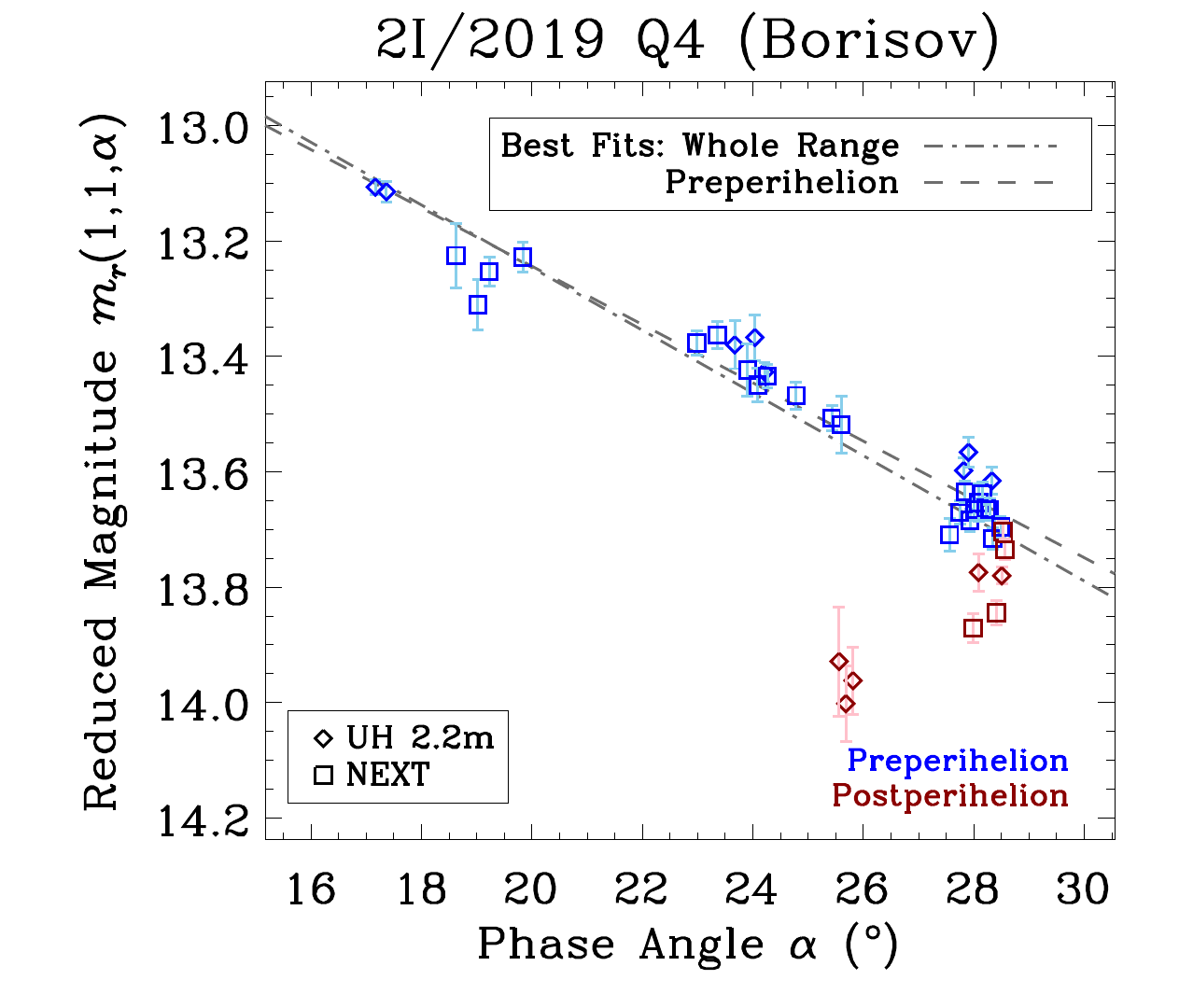}
\caption{
The reduced {\it r}-band magnitude of interstellar comet 2I/2019 Q4 (Borisov), $m_{r} \left(1,1,\alpha \right)$, versus the phase angle $\alpha$. Datapoints from the two observatories are plotted as different symbols, as indicated in the legend in the upper left corner of the figure. Preperihelion and postperihelion datapoints are in blue and dark red, respectively. The grey dashed line is the best linear least-squared fit to all of the {\it r}-band data, and the the less steep dashed dotted line is the best fit to the preperihelion counterparts (see Table \ref{tab:phi}).
\label{fig:2I_phi}
} 
\end{center} 
\end{figure}

\subsection{Phase Function}
\label{ss_phi}

When calculating the intrinsic brightness of 2I, we realised the actual phase function of the comet has a predominant influence on the result, as the phase angle varied nontrivially during the time period of our observing campaign. Thus, we feel the importance and necessity to discuss the phase function of 2I here.

In Figure \ref{fig:2I_phi}, we plot the reduced {\it r}-band magnitude of 2I (denoted as $m_{r}\left(1,1,\alpha\right)$, see its definition in Equation (\ref{eq_m_abs})) versus the phase angle, from which we can see that the datapoints from our earliest observations of the comet to those at maximum phase angle $\alpha \approx 30\degr$ (roughly corresponding to preperihelion epochs) appear to vary linearly with the phase angle. However, the trend for the datapoints starting from mid-December 2019 when the phase angle began to decrease is totally in disagreement with the earlier trend, indicating a change in the postperihelion activity of the comet. The obtained best-fit linear phase coefficient using the {\it r}-band magnitude datapoints from the whole timespan and the goodness of the fit are given in Table \ref{tab:phi}, where we can see that the reduced chi-square value is $\gg1$, suggesting an exceptionally poor fit. If we only fit the preperihelion part, however, the goodness of the fit is much more improved. Nevertheless, the best-fit phase coefficient, $\beta_{\alpha} = 0.0505 \pm 0.0010$ mag deg$^{-1}$, is larger than many (if not all) of the known solar system comets \citep[$0.02 \la \beta_{\alpha} \la 0.04$ mag deg$^{-1}$;][]{1987A&A...187..585M, 2019MNRAS.482.2924B}. We note that a considerably steeper backscattering phase function of the near-nucleus coma of comet 67P/Churyumov-Gerasimenko at small phase angles from the Rosetta mission was reported by \citet{2018Icar..309..265F}, who attribute the phenomenon to the presence of large transparent particles (at least \micron~size, and the imaginary index of refraction $\la$10$^{-2}$) from the nucleus in the region. Using the best-fit parameters and including the measurement uncertainties by \citet{2018Icar..309..265F}, we obtained that the $3\sigma$ upper limit to the phase coefficient in the same phase angle range as the one during our observed time period of 2I is $\beta_{\alpha} = 0.042$ mag deg$^{-1}$, which is inconsistent with the best-fit phase coefficient value we found for 2I (Table \ref{tab:phi}) at the $3\sigma$ level. By no means can we completely rule out the possibility that 2I has a phase function steeper than any of the known solar system comets. This scenario would require that the preperihelion activity of the comet would remain nearly constant, followed by a decline in the postperihelion activity. However, given the similarities between 2I and solar system comets, we prefer that the decline in the intrinsic brightness of 2I shown in Figure \ref{fig:2I_lc}b is due to the change in its activity.

\begin{deluxetable}{ccc}
\tablecaption{Best-Fit Phase Coefficients of Interstellar Comet 2I/2019 Q4 (Borisov)
\label{tab:phi}}
\tablewidth{0pt}
\tablehead{\colhead{Fitted Range} & \colhead{Phase Coefficient} & \colhead{Reduced Chi-Square\tablenotemark{\dag}} \\
& $\beta_{\alpha}$ (mag deg$^{-1}$) & $\chi_{\nu}^{2}$ }
\startdata
Whole & $0.0543 \pm 0.0009$ & 9.39\\
Preperihelion & $0.0505 \pm 0.0010$ & 2.47\\
\enddata
\tablenotetext{\dag}{Chi-square per degree of freedom, dimensionless.}
\tablecomments{Only the {\it r}-band datapoints were used to compute the best linear least-squared fits. See Figure \ref{fig:2I_phi} for the plot showing comparison between the best fits versus the data.}
\end{deluxetable}

\begin{figure*}
\gridline{\fig{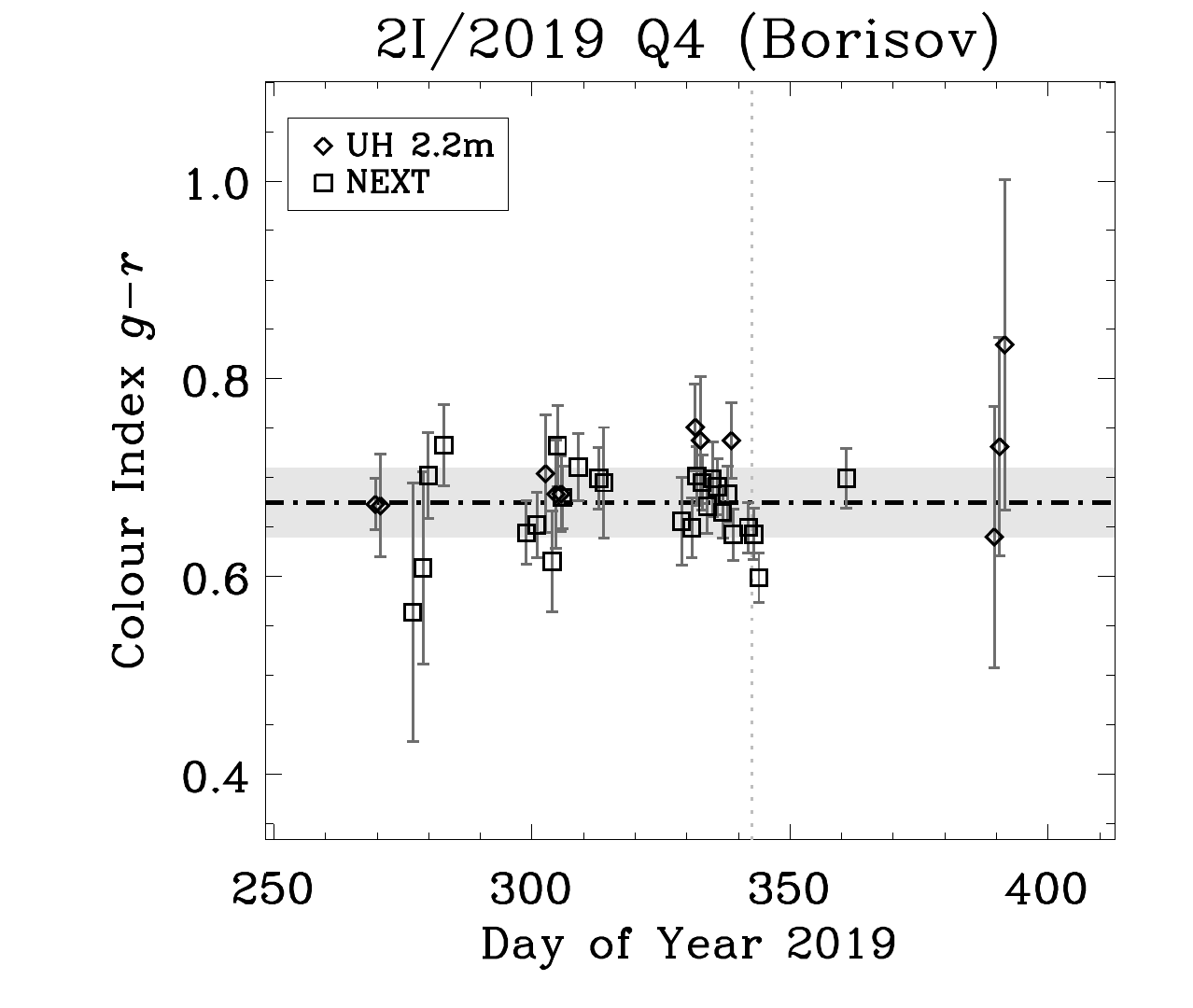}{0.5\textwidth}{(a)}
\fig{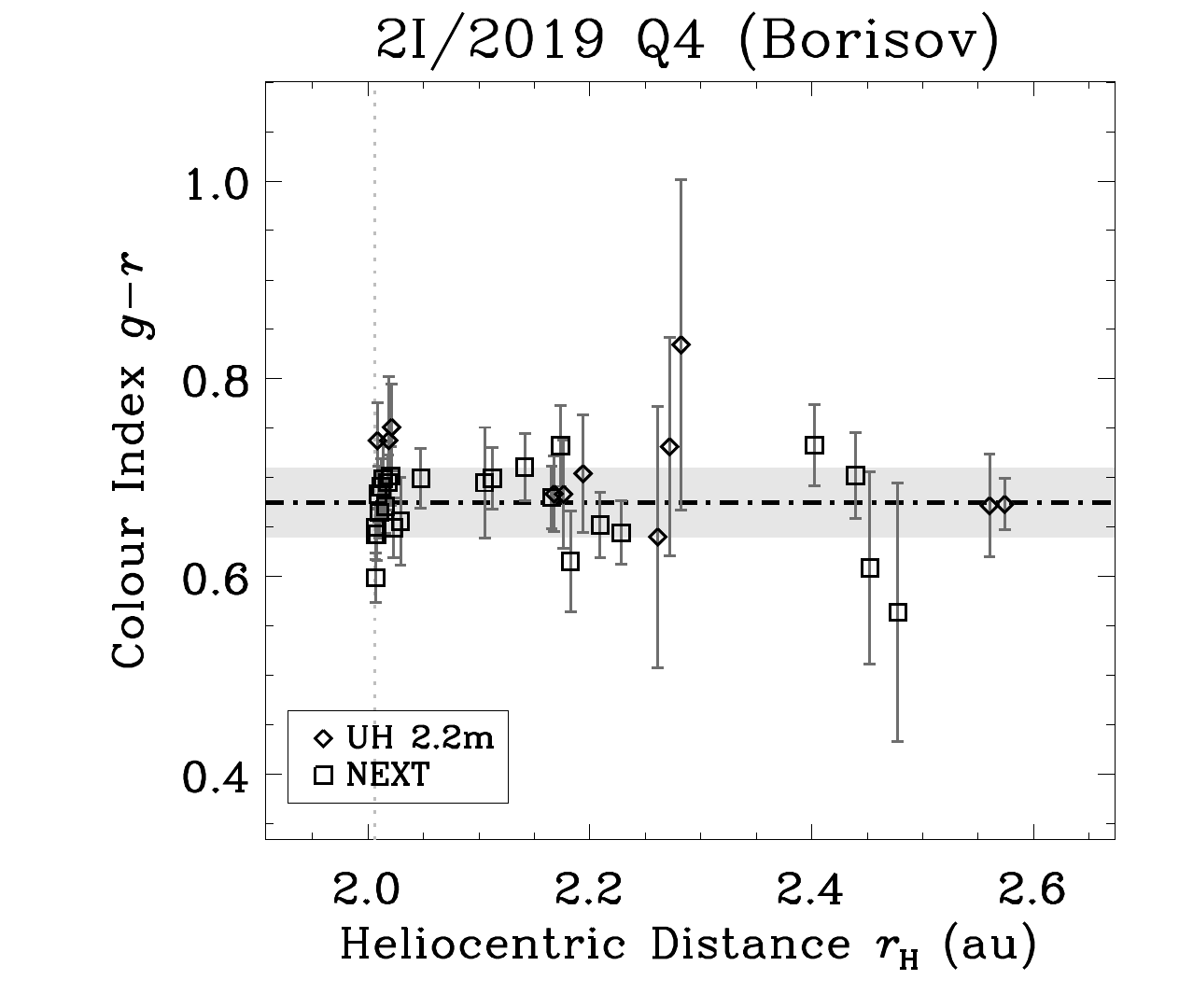}{0.5\textwidth}{(b)}}
\gridline{\fig{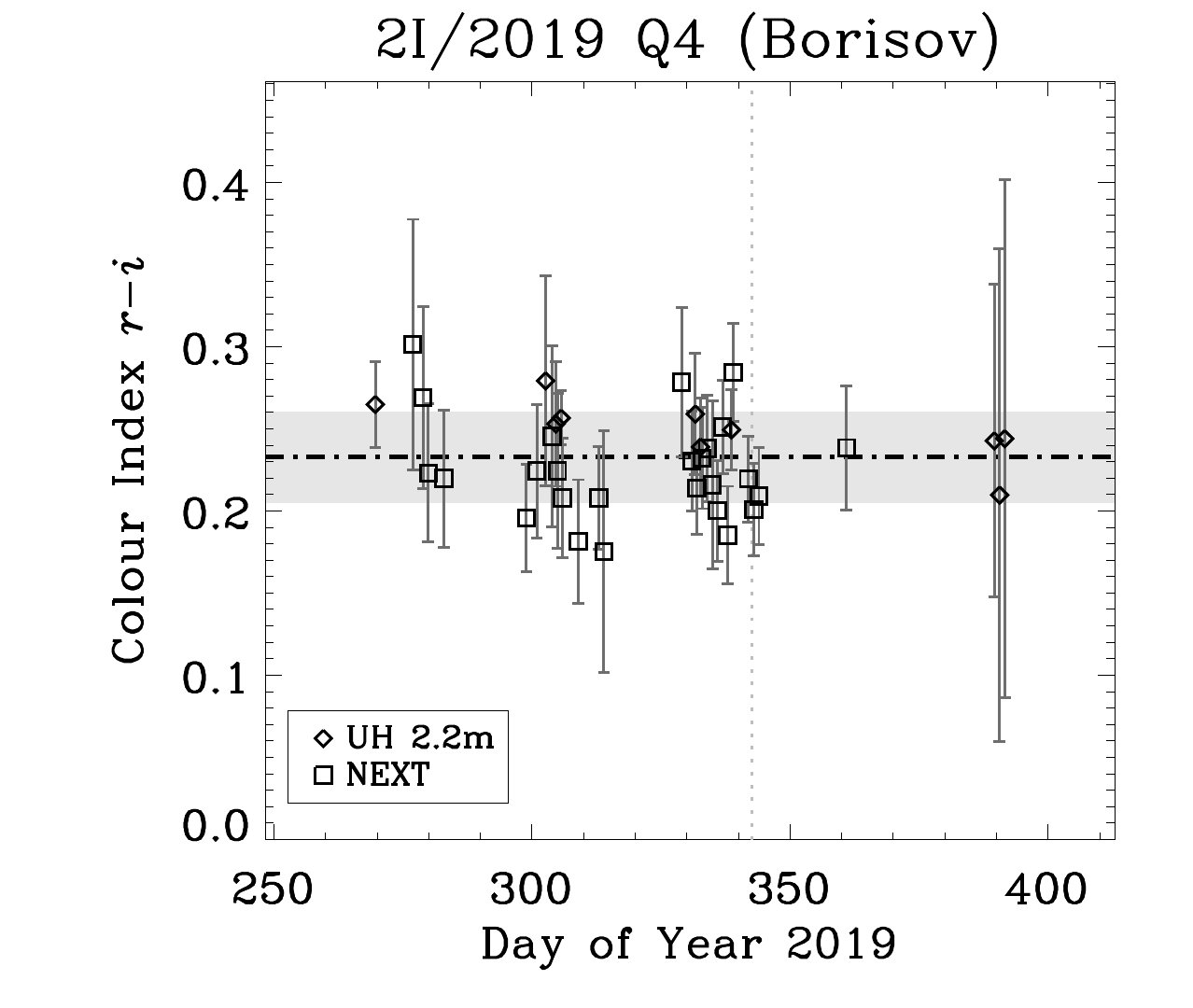}{0.5\textwidth}{(c)}
\fig{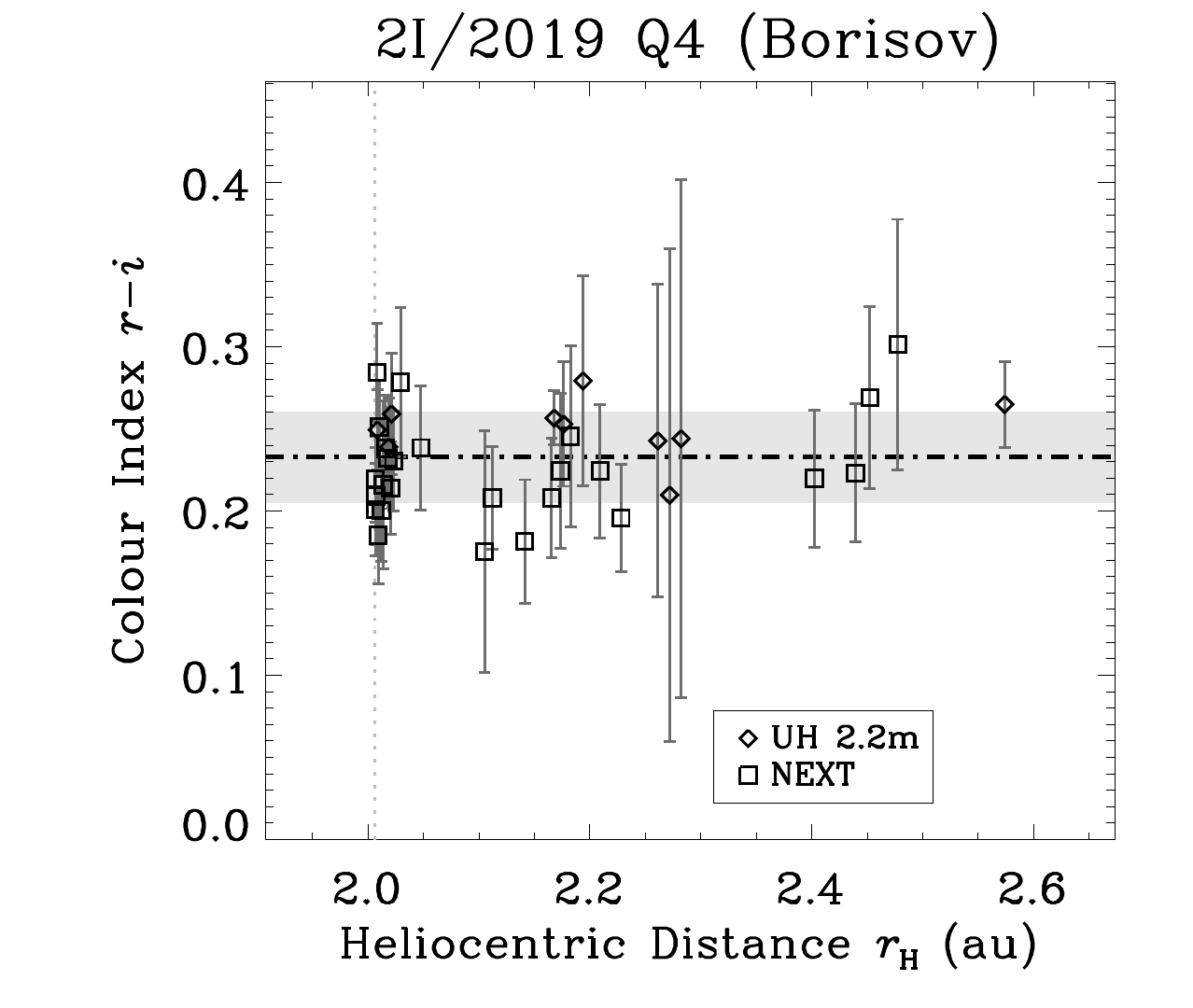}{0.5\textwidth}{(d)}}
\caption{
The evolution of the color indices {\it g} $-$ {\it r} (upper two panels) and {\it r} $-$ {\it i} (lower two panels) of interstellar comet 2I/2019 Q4 (Borisov) with time (left two panels) and the heliocentric distance (right two panels). Datapoints from the two observatories are discriminated by different point symbols, as indicated in the legends. Taking the measurement uncertainties into consideration, we see no evidence of colour variation of the comet. The time-average values of the colour indices are represented by a dashed-dotted horizontal line, with the grey zone labelling the $\pm1\sigma$ uncertainty region, in each of the panels. The perihelion epoch and distance of the comet are labelled as vertical grey dotted lines.
\label{fig:2I_clr}
} 
\end{figure*}

\begin{figure}
\begin{center}
\gridline{\fig{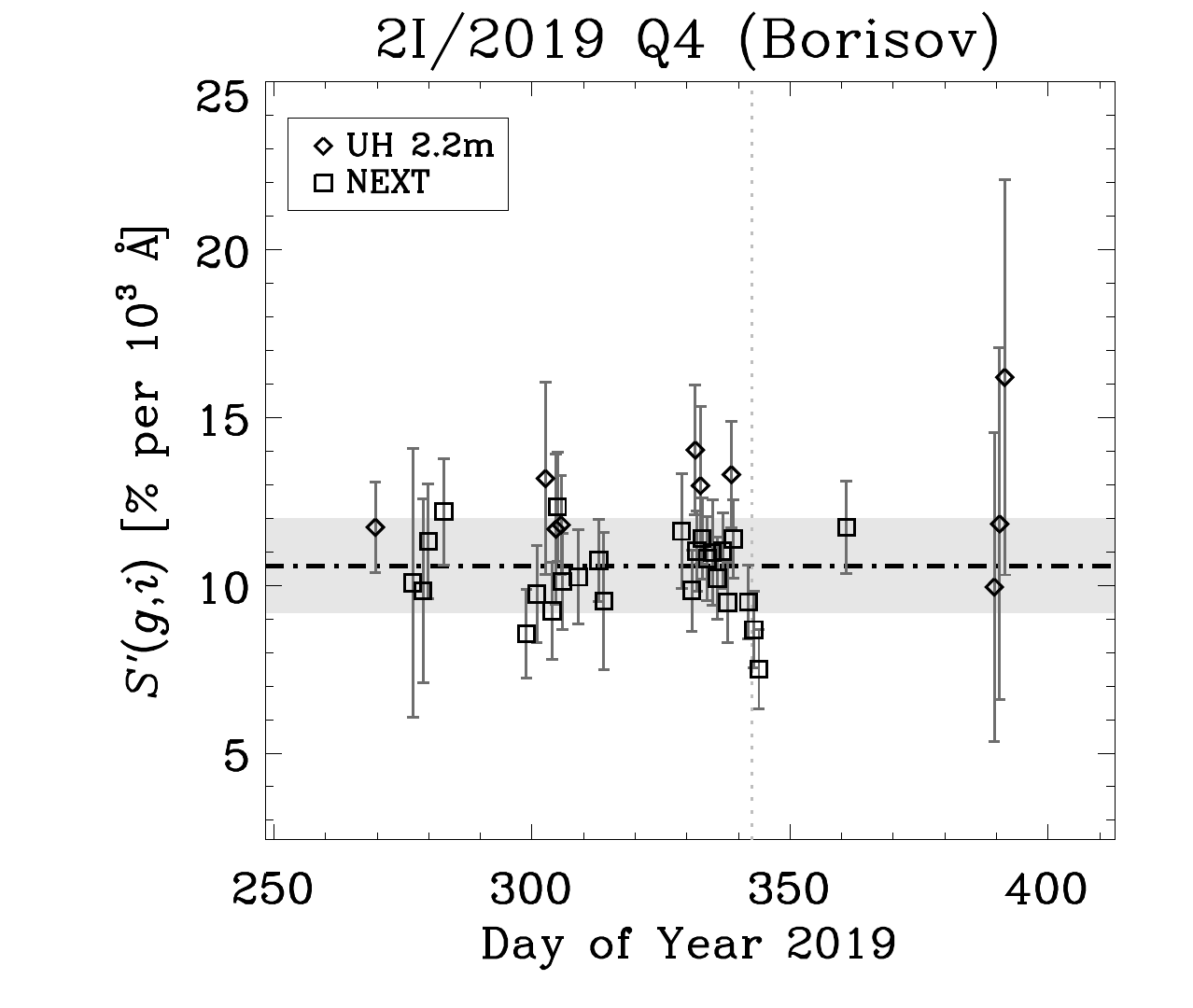}{0.45\textwidth}{(a)}}
\gridline{\fig{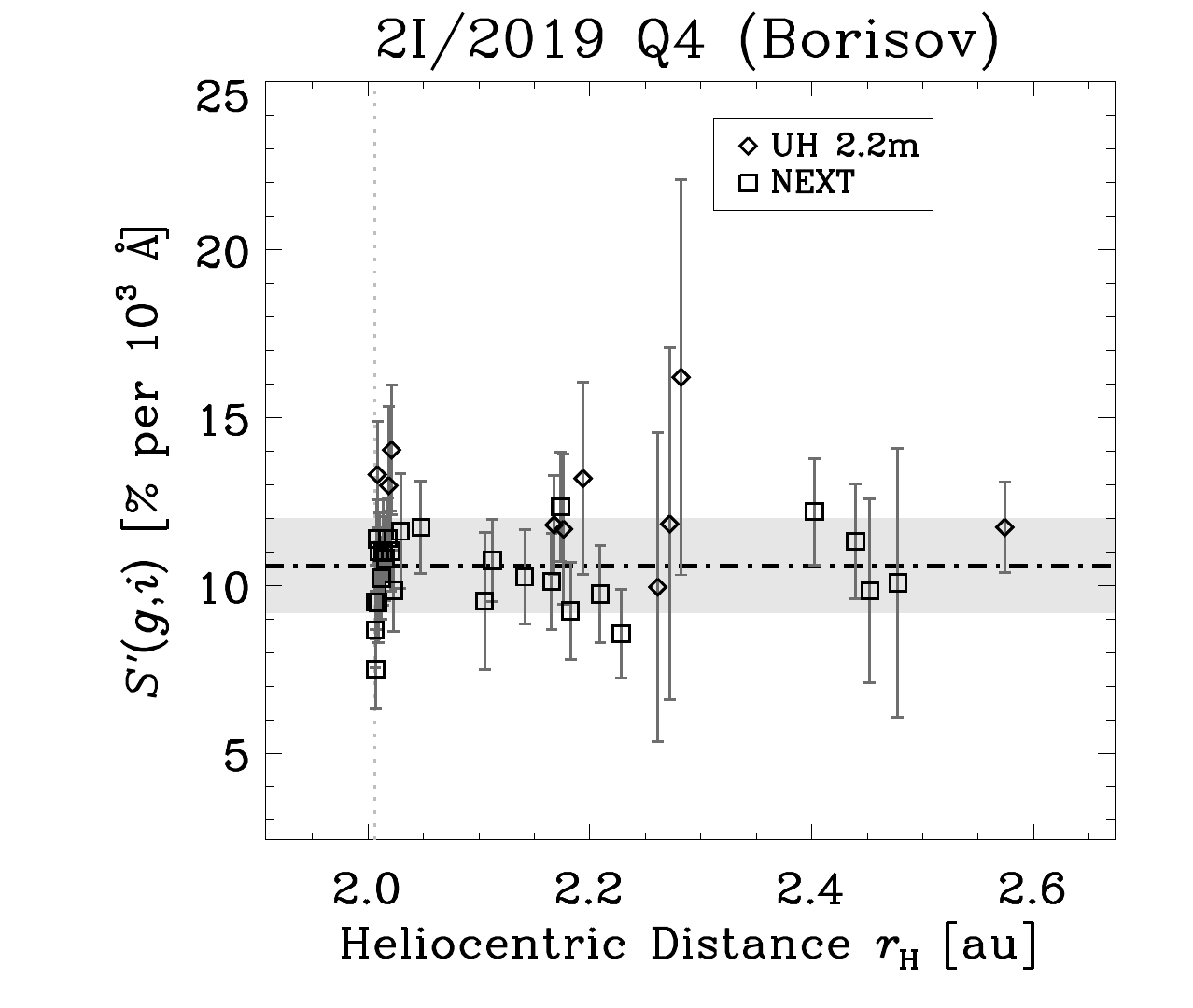}{0.45\textwidth}{(b)}}
\caption{
The normalised reflectivity gradient of interstellar comet 2I/2019 Q4 (Borisov) across the {\it g} and {\it i} bands as functions of time (a) and the heliocentric distance (b). No significant variation is seen. The horizontal dashed-dotted line is the mean value of the normalised reflectivity gradient, with the grey zone representing the associated $\pm1\sigma$ uncertainty. The vertical grey dotted lines in panels (a) and (b) mark the perihelion epoch and distance of the comet, respectively.
\label{fig:2I_sp}
} 
\end{center}
\end{figure}

\subsection{Color}
\label{ss_clr}

We plot the colour indices {\it g} $-$ {\it r} and {\it r} $-$ {\it i} of comet 2I versus both time and the heliocentric distance in Figure \ref{fig:2I_clr}. The transformation by \citet{Jordi06} was applied to derive the colour of the comet in the Johnson-Cousin system to the SDSS system. Generally speaking, similar to the solar system comets, the colour of 2I was slightly redder than the colour of the Sun \citep[$\left(g - r\right)_{\odot} = +0.46 \pm 0.02$, $\left(r - i\right)_{\odot} = +0.12 \pm 0.02$;][]{2018ApJS..236...47W} during our observing campaign from the UH 2.2 m telescope and NEXT. Given the uncertainties, we cannot spot any confident change in the colour of the comet, and thus derive the mean values of the colour indices as $\left \langle g - r \right \rangle = +0.68 \pm 0.04$, and $\left \langle r - i \right \rangle = +0.23 \pm 0.03$. The uncertainties are weighted standard deviation of the datapoints. Overall, the mean colour we found for the comet is consistent with the measurements by other observers \citep[e.g.,][]{2019arXiv191014004B,2019NatAs.tmp..467G,2019ApJ...886L..29J,2019A&A...631L...8O} if transformed to the same photometric system when necessary.

For completeness we also calculate the normalised reflectivity gradient \citep{1984AJ.....89..579A,1986ApJ...310..937J} across the {\it g} and {\it i} bands of the comet through the following equation:
\begin{equation}
S'\left(g, i \right) = \left( \frac{2}{\lambda_{i} - \lambda_{g}} \right) \frac{10^{0.4 \left[ \left(g-i\right) - \left(g-i\right)_{(\odot)}\right] } - 1}{10^{0.4 \left[ \left(g-i\right) - \left(g-i\right)_{(\odot)} \right] } + 1}
\label{eq_sp},
\end{equation}
\noindent where  $\lambda_g$ and $\lambda_i$ are the effective wavelengths of the {\it g} and {\it r} bands, respectively. A negative value of the normalised reflectivity gradient indicates that the colour of the object over the filter pair region is bluer than that of the Sun. Otherwise it is redder. We plot the results versus time and the heliocentric distance in Figure \ref{fig:2I_sp}. The uncertainties are propagated from the errors in the photometric measurements. Again, given the uncertainties of the datapoints we cannot identify any changes in the normalised reflectivity gradient of the comet. We obtain the mean value as $\overline{S'} \left(g,i\right) = \left( 10.6 \pm 1.4 \right) \%$ per $10^3$ \AA, which is in agreement with \citet{2019RNAAS...3..131D}, and is by no means outstanding in the context of known solar system comets and asteroids in cometary orbits \citep[e.g.,][]{2018A&A...618A.170L}. However, we are aware that differences in chemical composition between 2I and typical solar system comets have been noticed in various observations \citep{2020arXiv200111605B, 2020ApJ...889L..38K,2020NatAs.tmp...85B, 2020NatAs.tmp...84C}, the most marked of which is the unusually high CO abundance of 2I. The reason why we see no peculiarity in the broadband colour of the comet is likely that the signal we received was dominantly from scattering of sunlight by dust in the coma, rather than from gas emission.

\subsection{Morphology}
\label{ss_morph}

In the morphology of the dust tail of 2I lies the key to physical properties of the cometary dust grains therein. Their trajectories are determined by the initial ejection velocity ${\bf v}_{\rm ej}$, the release epoch, and the $\beta$ parameter, which is the ratio between the solar radiation pressure force and the gravitational force of the Sun, and is related to physical properties of dust grains by
\begin{equation}
\beta = \frac{\mathcal{C}_{\rm pr}\mathcal{Q}_{\rm pr}}{\rho_{\mathrm{d}} \mathfrak{a}}
\label{eq_beta}.
\end{equation}
\noindent Here, $\mathcal{C}_{\rm pr} = 5.95 \times 10^{-4}$ kg m$^{-2}$ is the solar radiation pressure coefficient, $\mathcal{Q}_{\rm pr} \approx 1$ is the scattering efficiency, $\mathfrak{a}$ and $\rho_{\rm d}$ are respectively the radius and the bulk density of the dust grains, assumed to be spherical. As the bulk density of dust grains in the coma of 2I remains unconstrained, we simply assume a constant value of $\rho_{\mathrm{d}} = 0.5$ g cm$^{-3}$, similar to the bulk density of typical solar system cometary nuclei \citep[e.g.,][]{2016Natur.530...63P}. The syndyne-synchrone computation \citep[e.g.,][]{1968ApJ...154..327F} by \citet{2020ApJ...888L..23J} suggests that the observed optically dominant dust grains of 2I are of $\beta \sim 0.01$, equivalent to a dust radius of $\mathfrak{a} \sim 0.1$ mm given our assumed value of the dust bulk density. However, a shortcoming of the syndyne-synchrone computation is that it ignores the initial ejection velocity of dust ${\bf v}_{\rm ej}$, which is physically unrealistic. A crude estimate of the ejection speed of the optically dominant dust grains can be gleaned by measuring the apparent length of the sunward extent to the dust coma of the comet from the following equation:
\begin{equation}
\left| {\bf v}_{\rm ej} \right| = \frac{\sqrt{2 \beta \mu_{\odot} {\it \Delta} \tan \ell \sin \alpha}}{r_{\rm H}}
\label{eq_vej_0},
\end{equation}
\noindent in which $\mu_{\odot} = 3.96 \times 10^{-14}$ au$^{3}$ s$^{-2}$ is the heliocentric gravitational constant, and $\ell$ is the apparent sunward turnaround angular distance. Our observations show $\ell \approx 2\arcsec$ in September 2019 to $\sim$3\arcsec~in December 2019. By inserting numbers into Equation (\ref{eq_vej_0}), we find that the ejection speed varied from $\left| {\bf v}_{\rm ej} \right| \approx 5$ m s$^{-1}$ in September 2019 to $\sim$8 m s$^{-1}$ around perihelion for the optically dominant dust grains of $\mathfrak{a} \sim 0.1$ mm in radius. If scaled to dust grains of $\beta \sim 1$, the ejection speed in September 2019 is in good line with the result by \citet[][$\left| {\bf v}_{\rm ej} \right| = 44 \pm 14$ m s$^{-1}$]{2019NatAs.tmp..467G}.

In order for us to better understand the morphology of the dust tail of 2I, we employ a more realistic Monte Carlo cometary ejection dust model to generate synthetic images of comet 2I. Except for the aforementioned parameters, the brightness profile of the dust coma is also governed by the size distribution of the dust grains, despite to a lesser extent. To maintain consistency with activity driven by volatile sublimation, we follow previous literature \citep[e.g.,][]{1950ApJ...111..375W,2008Icar..193...96I} and parameterise the dust ejection speed as
\begin{equation}
\left| {\bf v}_{\rm ej} \right| = V_0 \left( \frac{\beta}{r_{\rm H}} \right)^{1/2}
\label{eq_vej},
\end{equation}
\noindent where $V_0$ is the referenced ejection speed for dust grains with $\beta = 1$ ($\sim$1 \micron~in radius) at a heliocentric distance of $r_{\rm H} = 1$ au, and assume a simplistic power-law distribution for the dust size, with a fixed differential power-law index value of $\gamma = -3.6$ \citep{2004come.book..565F, 2019NatAs.tmp..467G}. This choice was made because our trial simulation shows that the spatial resolution and the SNR of our images are not sufficient to effectively constrain $\gamma$.

In our synthetic models, dust grains are released from the earliest observation of 2I by the ZTF in 2018 December \citep{2020AJ....159...77Y}, following a dust production rate $\sim r_{\rm H}^{-1}$ to maintain the consistency with the shallowness of the heliocentric distance dependency for the nongravitational effect of the comet (Section \ref{ss_NG}). We use the MERCURY6 package \citep{1999MNRAS.304..793C} to integrate dust grains of various values of $\beta$ and the release epochs to the observed epochs, taking gravitational perturbation from the major planets in the solar system into account, although this effect is trivial as the comet has no close encounters with any of the major planets. The Lorentz force is neglected because of its unimportance at the covered heliocentric distances of 2I \citep{2019AJ....157..103J, 2019AJ....157..162H}. The tridimensional distribution of the dust grains is then projected onto the sky plane viewed from Earth at some observed epoch. Thereby a bidimensional model image of the comet is formed, which is further convolved with a bidimensional Gaussian function with FWHM equal to the average FWHM of field stars in actual images, so as to mimic observational effects, including the instrumental point-spread effect and atmospheric seeing. The model image is then compared to the actual observations to identify the ranges of $V_0$ and $\mathfrak{a}_{\min}$ that can minimise discrepancies between the two sets of images in the least-square space. Expectedly and through testing, we cannot constrain the maximum size of the dust grains from our observations, as they do not travel afar from the nucleus. We thus somewhat arbitrarily adopt $\mathfrak{a}_{\max} = 0.1$ m as a fixed parameter.

\begin{figure}
\epsscale{1.2}
\begin{center}
\plotone{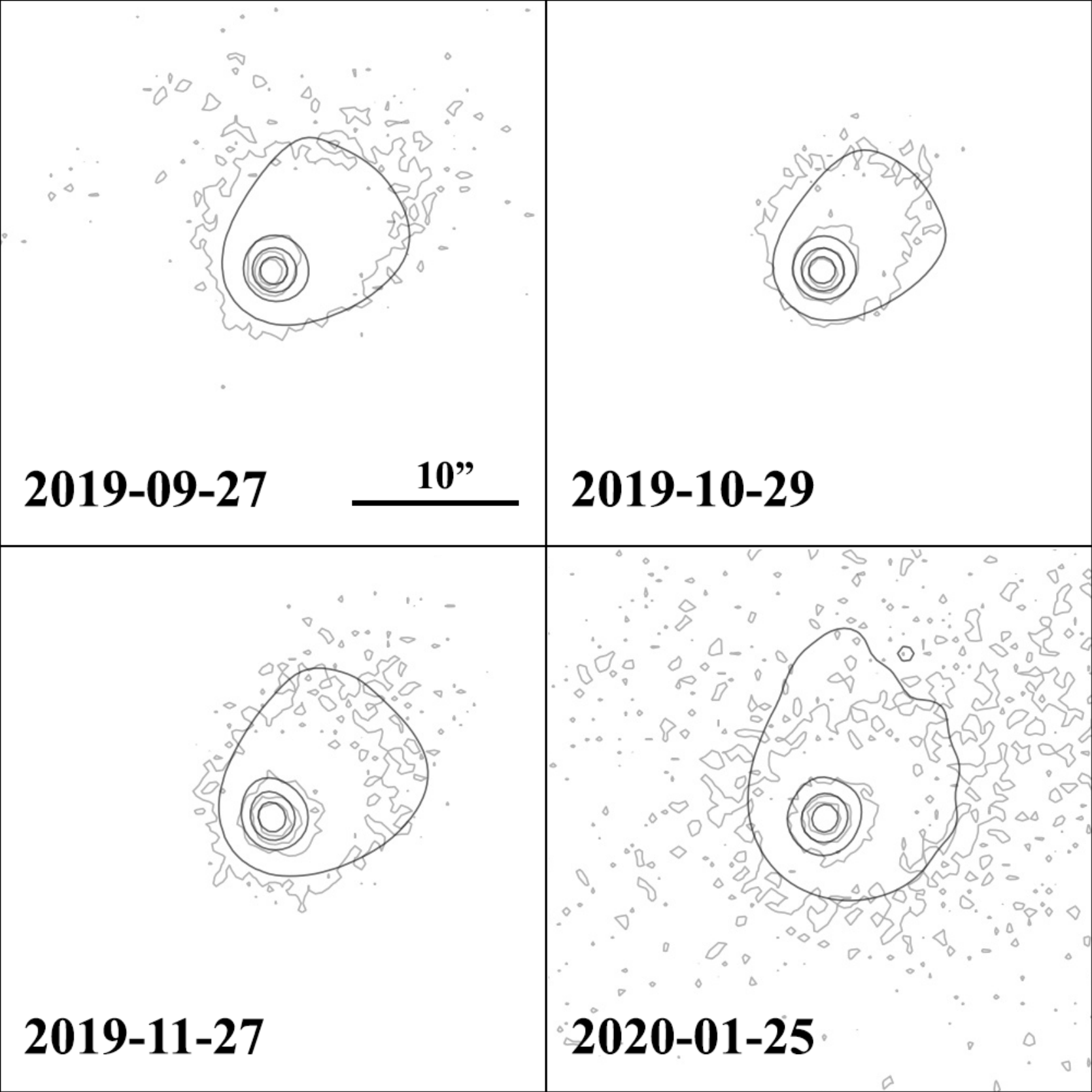}
\caption{
Comparison between the best-fit modelled (white contours) and observed (background images) morphology of interstellar comet 2I/2019 Q4 (Borisov) on selected dates. A scale bar of 10\arcsec~in length, applicable to all of the four panels, is shown. Equatorial north is up and east is left. See Figure \ref{img:2I_UH88} for the position angles of the antisolar direction and the negative heliocentric velocity projected onto the sky plane. The best-fit parameters are listed in Table \ref{tab:mdl}.
\label{fig:mdl}
} 
\end{center} 
\end{figure}

We summarise the best-fit results in Table~\ref{tab:mdl}, with the best-fit models in comparison with observed image shown in Figure~\ref{fig:mdl}. Although the assumption of the dust production rate following $\sim r_{\rm H}^{-1}$ was made to obtain the best-fit results, we found that the results are relatively insensitive to the assumption. For instance, if we instead assume a steeper heliocentric distance dependency for the dust production rate $\sim r_{\rm H}^{-2}$, the best-fit results remain completely the same. By applying Equation (\ref{eq_vej}), we can see that the best-fit ejection speed for the optically dominant dust grains in the tail of the comet is in great agreement with the crude estimate from the sunward turnaround point. Our result for the UH 2.2 m observation from 2019 September 27 is consistent with \citet[][$V_{0} = 74 \pm 23$ m s$^{-1}$]{2019NatAs.tmp..467G}. So is our result for October 2019, if compared to \citet[$\sim$90 m s$^{-1}$]{2020arXiv200502468K}. Thus, we are confident to conclude that the dust ejection speeds of the comet appear to be low, only $\sim$15-30\% of the ejection speed given by an ejection model assuming sublimation of water ice. In this regard, 2I is similar to some of the low-activity comets in the solar system, such as 209P/LINEAR \citep{2016Icar..264...48Y}. Meanwhile, we notice that $V_{0}$ possibly has been increasing during our observed period of the comet. Similar phenomena amongst solar system comets have been identified observationally \citep[e.g.,][]{2011A&A...531A..54T, 2014ApJ...791..118M}. Yet given the large uncertainty we opt not to further interpret it.

Our best-fit results also indicate that the minimum grain size in the dust tail of 2I, which is found to be in a range between $\sim$3 \micron~and 1 mm in radius, is unsurprising in the context of solar system comets \citep[e.g.,][]{2004come.book..565F}. We thereby infer that the activity mechanism on 2I likely resembles that of typical comets in the solar system, whose dust grains are ejected from the nucleus surface by coupling with the gas flow of outgassing volatiles. 

\begin{deluxetable}{c|cc}
\tablecaption{Best-Fit Parameters for Dust Coma Morphology Modelling of Comet 2I/2019 Q4 (Borisov)
\label{tab:mdl}
}
\tablewidth{0pt}
\tablehead{Time & Referenced Ejection Speed & Minimum Radius \\
(UT) & $V_{0}$ (m s$^{-1}$) & $\mathfrak{a}_{\min}$ (m) }
\startdata
2019 Sep 27 & $70\pm10$ & $10^{-4.5 \pm 0.5}$ \\
2019 Oct 29 & $80\pm20$ & $10^{-4.0 \pm 0.5}$ \\
2019 Nov 27 & $100\pm20$ & $10^{-4.0 \pm 1.0}$ \\
2020 Jan 25 & $140\pm20$ & $10^{-5.0 \pm 0.5}$ \\
\enddata
\tablecomments{We test the referenced dust ejection speed $V_{0}$ in a step size of 10 m s$^{-1}$ for $V_{0} \in \left[0, 400\right]$ m s$^{-1}$, and the minimum dust radius $\mathfrak{a}_{\min} \in \left \{10^{\left \lfloor  \log \mathfrak{a}_{\min} \right \rfloor / 2} \cap \left[10^{-6.5}, 10^{-3.0}\right] \right \}$ m.
}
\end{deluxetable}

\subsection{Activity}
\label{ss_act}

\begin{figure}
\epsscale{1.2}
\begin{center}
\plotone{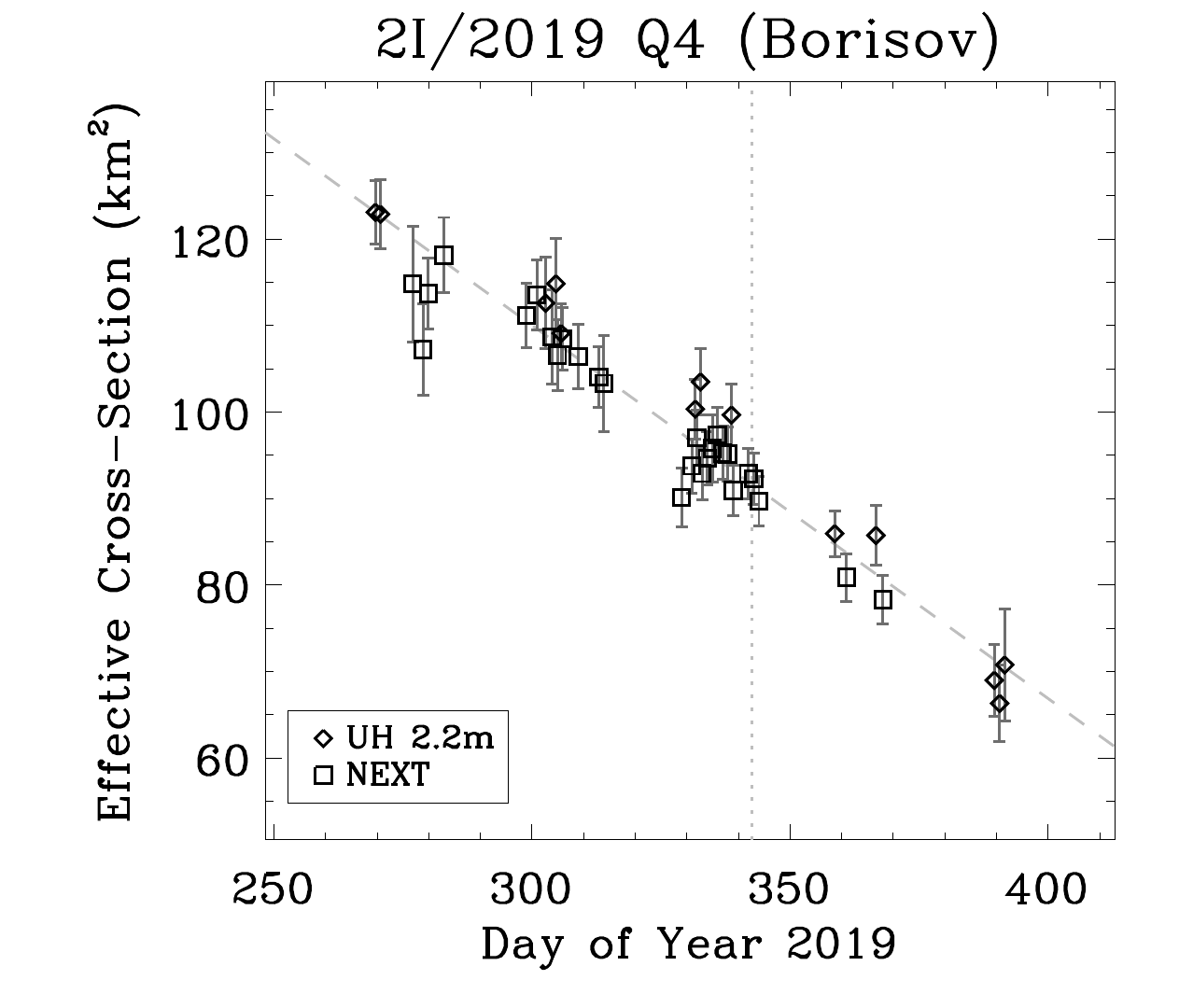}
\caption{
Temporal variation of the effective geometric scattering cross-section, converted from the {\it r}-band magnitude measurements, of interstellar comet 2I/2019 Q4 (Borisov). Apparently the downtrend is noticeable. Datapoints from the two observatories are discriminated by different point symbols as shown in the legend of the plot in the lower left corner. The grey dashed line is the best linear least-squared fit, whereas the vertical dotted grey line marks the perihelion epoch of the comet.
\label{fig:2I_XS}
} 
\end{center} 
\end{figure}

The activity level of the comet can be assessed through investigating its effective geometric scattering cross-section, which we compute from the {\it r}-band magnitude measurements using the following equation:
\begin{equation}
C_{\rm e} = \frac{\pi r_{0}^{2}}{p_{r}} 10^{0.4 \left[m_{\odot,r} - m_{r} \left(1,1,0 \right) \right]}
\label{eq_XS},
\end{equation}
\noindent where $C_{\rm e}$ is the effective scattering cross-section, $p_{r}$ is the geometric albedo of cometary dust in the coma of the comet, and $m_{\odot, r} = -26.93$ is the apparent {\it r}-band magnitude of the Sun at the mean Earth-Sun distance $r_{0} = 1.5 \times 10^{8}$ km \citep{2018ApJS..236...47W}. So far there is no constraint on the value of $p_{r}$ of comet 2I, and therefore we assume $p_{r} = 0.1$ as the appropriate value for cometary dust \citep[e.g.,][]{2017JQSRT.202..104Z}. The effective cross-section as a function of time of the comet is shown in Figure \ref{fig:2I_XS}, in which the decline of the intrinsic brightness of the comet is translated to a continuous decline in the effective cross-section within the photometric aperture of $\varrho = 10^{4}$ km in radius, which can be excellently approximated by a linear function with a best-fit slope of $\overline{\dot{C}_{\rm e}} = -0.43 \pm 0.02$ km$^{2}$ d$^{-1}$. We note that this result is incompatible with the observations by \citet{2019ApJ...886L..29J} and \citet{2020ApJ...888L..23J}. However, given the fact that our observation covered a much wider period range, we argue that there is likely short-term scale variability in the activity of the comet. We do not recall witnessing similar overall trends for the known solar system comets, whose effective scattering cross-sections generally increase and then decrease as they approach and recede from the Sun, respectively, unless outbursts occur. A plausible explanation is that the coma of 2I consists of an abundant number of icy grains that continuously sublimate until exhaustion of volatiles, whereby the effective scattering cross-section diminishes. There exists an alternative explanation that the preperihelion downtrend in Figure \ref{fig:2I_XS} is bogus if the actual phase function of 2I is unprecedentedly steeper than those of the known solar system comets (Section \ref{ss_phi}). Considering the similarities between 2I and solar system comets, we do not prefer this explanation.

The average net mass-loss rate in the fixed-size photometric aperture is related to the mean change rate of the effective cross-section by
\begin{equation}
\overline{\dot{M}_{\rm d}} = \frac{4}{3}\rho_{\rm d} \bar{\mathfrak{a}} \overline{\dot{C}_{\rm e}}
\label{eq_mloss}.
\end{equation}
\noindent Given the difficulty in determining the mean dust radius $\bar{\mathfrak{a}}$ as the maximum dust size cannot be well constrained, we instead use the optically dominant dust size $\beta \sim 0.01$. Thus we obtain that the mean net mass-loss rate of comet 2I is $\overline{\dot{M}_{\rm d}} \approx - 0.4 $ kg s$^{-1}$ over the whole observed period. The negative value means that the newly produced mass in the photometric aperture of $\varrho = 10^{4}$ km fails to supply the mass that leaves the aperture due to the the solar radiation pressure force and/or nonzero ejection speeds.

As we cannot really constrain the actual mass loss of comet 2I in the above manner, instead we examine it based upon our detection of the nongravitational effect of the comet, in essence owing to the momentum conservation:
\begin{equation}
\kappa \dot{M}_{\rm n} v_{\rm out} = -M_{\rm n} g\left( r_{\rm H} \right) \sqrt{\sum_{j=1}^{3} A_{j}^{2}}
\label{eq_mc}.
\end{equation}
Here, $0 \le \kappa \le 1$ is the collimation efficiency coefficient of mass ejection, with the lower and upper boundaries corresponding to isotropic and perfectly collimated ejection, respectively, $M_{\rm n}$ is the nucleus mass of the comet, $g\left( r_{\rm H} \right)$ is the adimensional nongravitational force function (tested in Section \ref{ss_NG} as different power laws $\sim r_{\rm H}^{-n}$) that follows the $r_{\rm H}$-dependency for the nongravitational acceleration and is normalised at $r_{\rm H} = 1$ au, and $v_{\rm out}$ is the outflow speed of mass-loss materials, which is approximated by the empirical function given in \citet{2004come.book..523C}. As mentioned in Section \ref{ss_NG}, $g\left( r_{\rm H} \right) \sim r_{\rm H}^{-1 \pm 1}$ is preferred by the fit to the astrometric data of 2I, and therefore we investigate scenarios with three different power-law indices $n=0$, 1 and 2 in the nongravitational force function. 

\begin{figure}
\epsscale{1.2}
\begin{center}
\plotone{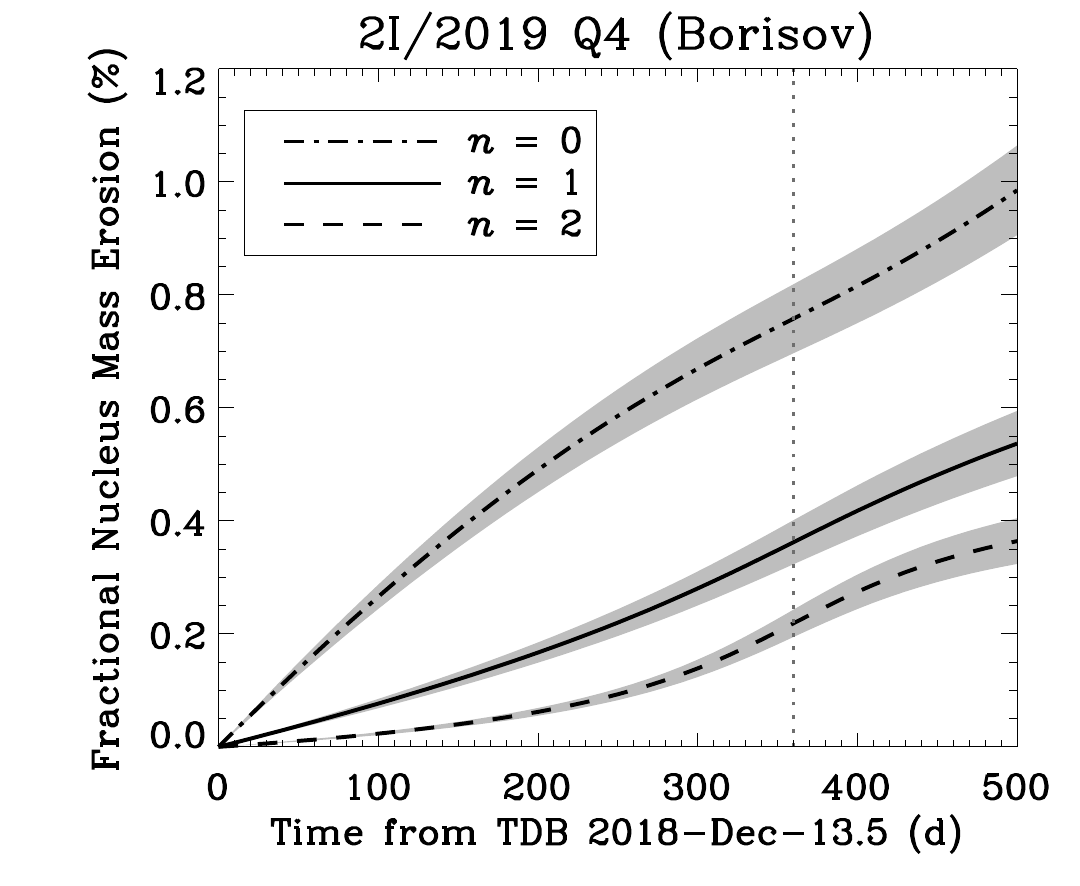}
\caption{
Fractional nucleus mass erosion of interstellar comet 2I/2019 Q4 (Borisov) as functions of time with different power-law indices $n$ in the nongravitational force function (discriminated by line styles, see the legend). The corresponding $1\sigma$ uncertainty regions, propagated from the formal errors in the nongravitational parameters, are shaded as grey zones. Here we assume $\kappa = 1$ for the collimation efficiency coefficient, i.e., perfectly collimated mass loss of the nucleus. The vertical dotted grey line is the perihelion epoch of the comet.
\label{fig:2I_MEros}
} 
\end{center} 
\end{figure}

Equation (\ref{eq_mc}) can be separable and integrable to find the expression for the fractional mass erosion of the comet between epochs $t_{1}$ and $t_{2}$:
\begin{align}
\nonumber
\mathcal{E}_{\rm M} \left(t_1, t_2 \right) & \equiv 1 - \frac{M_{\rm n} \left(t_2 \right)}{M_{\rm n} \left(t_1 \right)} \\
& \approx \frac{ r_{0}^{n} }{\kappa} \sqrt{\sum_{j=1}^{3} A_{j}^2} \int\limits_{t_{1}}^{t_{2}} \frac{{\rm d}t}{r_{\rm H}^{n} \left(t\right) v_{\rm out} \left( t \right)}
\label{eq_erosm}.
\end{align}
\noindent We take $t_{1} \approx $ TDB 2018 December 13.5, i.e., the earliest detection of the comet by the ZTF \citep{2020AJ....159...77Y}. The fractional mass erosion of the nucleus as functions of time following the three best-fit nongravitational force models (power-law indices $n=0,1$, and 2) with the assumption of a maximum collimation efficiency coefficient of mass ejection, i.e., $\kappa = 1$, corresponding to perfectly collimated ejection, is plotted in Figure \ref{fig:2I_MEros}. As expected, the nucleus of 2I will be eroded maximally in the case of $n = 0$, because the magnitude of its nongravitational acceleration is a constant, whereas the erosion will be the smallest in the inverse-square case, because the magnitude of the nongravitational acceleration drops the most steeply amongst the three cases as the heliocentric distance increases. Given the range of the collimation efficiency coefficient, we can conclude that since the earliest detection by the ZTF of the comet in mid-December 2019, $\ga$0.2\% of the total nucleus mass has been eroded. The erosion continued to increase as the comet passed perihelion. There is an additional uncertainty in the outflow speed that we have not included in the above calculation, and therefore we suggest that the results for the fractional mass erosion of the comet are better treated as order-of-magnitude estimates.

\subsection{Nucleus Size}
\label{ss_nucsz}

The detection of the nongravitational effect in the motion of 2I enables us to evaluate its nucleus radius. We transform Equation (\ref{eq_mc}) into
\begin{equation}
R_{\rm n} \approx \left[\frac{3 \kappa \mathscr{U} \mathfrak{m}_{\rm H} Q v_{\rm out}}{4 \pi \rho_{\rm n} \sqrt{\sum_{j=1}^{3} A_{j}^{2}}} \left(\frac{r_{\rm H}}{r_{0}} \right)^{n}  \right]^{1/3}
\label{eq_nucrad},
\end{equation}
\noindent where $\mathscr{U}$ and $Q$ are respectively the molecular weight and the production rate of the dominant mass-loss substance of the comet, and $\mathfrak{m}_{\rm H} = 1.66 \times 10^{-27}$ kg is the mass of the hydrogen atom. We assume a typical cometary nucleus density, $\rho_{\rm n} = 0.5$ g cm$^{-3}$ \citep[e.g.,][]{2016Natur.530...63P}. Since 2I is the most CO-rich in the context of solar system comets \citep{2020NatAs.tmp...85B, 2020NatAs.tmp...84C}, with the only exception of C/2016 R2 (PANSTARRS) \citep{2019AJ....158..128M}, it therefore seems more likely that the the detected nongravitational motion of the comet is caused by outflow of CO. We constrain the nucleus size using results from \citet{2020NatAs.tmp...84C}, who reported $Q_{\rm CO} = \left(4.4 \pm 0.7 \right) \times 10^{26}$ s$^{-1}$ at $r_{\rm H} = 2.01$ au with an outflow speed of $v_{\rm out} = \left(4.7 \pm 0.4\right) \times 10^{2}$ m s$^{-1}$. Substituting, we find $R_{\rm n} \la \left(3.6 \pm 0.3\right) \times 10^{2}$ m.

In fact, even if we ignore the fact that the activity of 2I is CO-dominant \citep{2020NatAs.tmp...85B, 2020NatAs.tmp...84C} but assume momentarily that the nongravitational effect is caused by sublimation of H$_2$O ice, there is still no major change to the upper limit to the nucleus size of 2I. While multiple observers have reported emission lines of the comet \citep[e.g.,][]{2019ApJ...885L...9F, 2019A&A...631L...8O, 2020ApJ...889L..10M,2020ApJ...889L..38K}, the most straightforward estimate of the mass-loss rate of the comet is constrained from the detection of the forbidden oxygen ([O\,{\footnotesize I}] 6300 \AA) line by \citet{2020ApJ...889L..10M}, who assumed that H$_{2}$O is the dominant source and derived a production rate of $Q_{\rm H_{2}O} = \left(6.3 \pm 1.5 \right) \times 10^{26}$ s$^{-1}$ at heliocentric distance $r_{\rm H} = 2.38$ au. Plugging numbers in, Equation (\ref{eq_nucrad}) yields an upper limit to the nucleus radius of 2I as $R_{\rm n} \la \left(3.9 \pm 0.5 \right) \times 10^{2}$ m, consistent with the estimate using measurements from \citet{2020NatAs.tmp...84C}. 

We are therefore confident to conclude that, based on our detection of the nongravitational acceleration of 2I, the nucleus is most likely $\la$0.4 km in radius, in excellent agreement with the HST observation by \citet{2020ApJ...888L..23J}.

\section{Summary}
\label{sec_sum}

We summarise our study of the observations from the UH 2.2 m telescope and 0.6 m NEXT of interstellar comet 2I/2019 Q4 (Borisov) in the following.

\begin{enumerate}

\item The intrinsic brightness of the comet was observed to decline starting from late September 2019 to late January 2020, on its way from preperihelion all the way to the outbound leg postperihelion. This behaviour, which appears uncommon in the context of solar system comets without outbursts, is likely unexplained by the phase effect but the downtrend of the effective scattering cross-section due to sublimation of volatiles with a slope of $-0.43 \pm 0.02$ km$^{2}$ d$^{-1}$.

\item We have a statistically confident detection of the nongravitational acceleration of the comet, which follows a shallow heliocentric distance dependency of $r_{\rm H}^{-1 \pm 1}$, with the available astrometric observations. By perihelion, a fraction of $\ga$0.2\% of the total nucleus in mass has been eroded since the earliest detection by the ZTF in mid-December 2018.

\item Assuming a typical cometary nucleus density ($\rho_{\rm n} = 0.5$ g cm$^{-3}$), we estimate from the detected nongravitational effect of the comet that its nucleus is most likely $\la$0.4 km in radius, in favour of the result from the HST observation by \citet{2020ApJ...888L..23J}.

\item Our morphologic analysis of the dust tail reveals that the ejection speed increased from $\sim$4 m s$^{-1}$ in September 2019 to $\sim$7 m s$^{-1}$ for the optically dominant dust grains of $\beta \sim 0.01$ (corresponding to a grain radius of $\mathfrak{a} \sim 0.1$ mm, given an assumed dust bulk density of $\rho_{\rm d} = 0.5$ g cm$^{-3}$). The dust grains with contribution to the effective geometric scattering cross-section are no smaller than micron size.

\item The colour of the comet remained unchanged with the uncertainty taken into consideration, which is, on average, unexceptional in the context of known solar system comets. We determined the mean values of the colour as $\left \langle g - r \right \rangle = 0.68 \pm 0.04$, $\left \langle r - i \right \rangle = 0.23 \pm 0.03$, and the normalised reflectivity gradient over the {\it g} and {\it i} bands $\overline{S'} \left(g,i\right) = \left(10.6 \pm 1.4\right)$ \% per $10^3$ \AA.

\end{enumerate}

\acknowledgements
{
We thank Luke McKay and the engineer team of the UH 2.2 m telescope, as well as Xing Gao of Xingming Observatory, for technical assists. We also thank Davide Farnocchia, Bill Gray, David Jewitt, Yoonyoung Kim, and Adam McKay for insightful help and discussions, and the observers who submitted good-quality astrometric measurements of the comet to the MPC. Comments from an anonymous reviewer greatly help us improve the quality of this work. The research was funded by NASA Near Earth Object Observations grant No. NNX13AI64G to D.J.T.
}

\vspace{5mm}
\facilities{0.6 m NEXT, UH 2.2m.}

\software{FindOrb, HOTPANTS \citep{2015ascl.soft04004B}, IDL, IRAF, L.A.Cosmic \citep{2001PASP..113.1420V}, MERCURY6 \citep{1999MNRAS.304..793C}, StarFinder \citep{2000A&AS..147..335D}.}


\end{document}